\documentclass[%
 reprint,
 amsmath,amssymb,
 aps,
]{revtex4-2}
\usepackage{float}
\usepackage{xcolor, soul}
\sethlcolor{green}
\usepackage{ulem}
\newcommand\redout{\bgroup\markoverwith
{\textcolor{red}{\rule[.5ex]{4pt}{1.2pt}}}\ULon}
\usepackage{siunitx}
\usepackage{graphicx}
\usepackage{dcolumn}
\usepackage{bm}
\usepackage[section]{placeins}
\usepackage{url}
\usepackage{mdframed}

\newcommand{\zit}[1]
       {\cite{#1}}  % choose one of both!
\usepackage{caption}
\begin{document}

\preprint{APS/123-QED}

\title{Signatures of ballistic and diffusive transport\\ in the time-dependent Kerr-response of magnetic materials} 
\author{Sanjay Ashok$^1$, Jonas Hoefer$^1$, Martin Stiehl$^1$, Martin Aeschlimann$^1$, Hans Christian Schneider$^1$, Baerbel Rethfeld$^1$,  Benjamin Stadtm\"uller$^{1,2}$}
 \affiliation{$^1$Department of Physics and OPTIMAS Research Center, RPTU Kaiserslautern-Landau, Erwin Schr\"odinger Stra{\ss}e 46, 67663 Kaiserslautern, Germany\\
 $^2$Experimentalphysik II, Institute of Physics, Augsburg University, 86159 Augsburg, Germany
 }

\date{\today}% It is always \today, today,
             %  but any date may be explicitly specified

\begin{abstract}

We calculate the influence of diffusive and ballistic transport  on ultrafast magnetization in thick metallic films. 
When only diffusive transport is present, gradients of magnetization in the material remain up to picosecond timescales. 
In contrast, in the extreme superdiffusive limit where ballistic transport dominates, the magnetization changes homogeneously in space. We calculate the measurable magneto-optical responses for a $\SI{40}{\nano\meter}$ Nickel film. Although the resulting Kerr rotation dynamics are found to be very similar in the two limits of transport, our simulations reveal a clear signature of magnetization gradients in the Kerr ellipticity dynamics, namely a strong probe-angle dependence for the case when diffusive transport allows gradients to persist. We then perform probe-angle dependent complex magneto-optical Kerr effect (CMOKE) measurements on an excited \SI{40}{\nano\meter} Nickel film. The angle dependence of the measured Kerr signals closely matches the simulated response with diffusive transport. Therefore we conclude that the influence of ballistic transport on ultrafast magnetization dynamics in such films is negligible.

\end{abstract}

\maketitle

\section{Introduction}

Spin currents are the fuel of magnetic or spintronic information technology devices \cite{Kent2015, Erve2009, Kampfrath2013, Hoffmann2015}. They mediate the transport of information encoded in the spin angular momentum of the electron and allow information to be written into magnetic storage devices \cite{Chappert2007, Lebrun2018}, for example by spin transfer or spin-orbit torques \cite{Pfeiffer2021, Reimers2023}. These concepts are now well established for electronically generated spin currents, which allow the manipulation of magnetic order on timescales down to a few hundred picoseconds \cite{Cheng2016, Rudolf2012, Melnikov2022, Choi2014}.
\par
{Faster manipulation of magnetic materials by spin transport can only be achieved by using optically generated spin currents }. These spin-polarized charge currents are typically generated during the ultrafast magnetization of ferromagnetic materials after optical excitation with fs light pulses \cite{Ferte2019}. After the first observation of ultrafast spin currents in magnetization experiments \cite{Beaurepaire1996}, most studies have focused on the investigation of the evolution of spin transport in magnetic bi- and multi-layer structures \cite{Dutta2022, Rudolf2012, Melnikov2011, Melnikov2022}. This has led, for example, to the demonstration of ultrafast spin transfer torques \cite{Schellekens2014, Chirac2020} and spin-current-driven magnetization reversal in magnetic multilayer structures \cite{Gerrits2002, Remy2023}. These studies have not only revealed the potential of spin currents to alter magnetic order on ultrafast timescales, but have also provided insights into ultrafast spin current generation and spin injection efficiency across interfaces. Much less is known about the evolution of ultrafast spin currents in a single layer or in homogeneous materials \cite{Krieger2017}. This is mainly due to the fact that it is experimentally extremely difficult to access the dynamics of optically induced spin currents in a single material in real time.
\par
The typical method of choice here is based on the magneto-optical Kerr effect (MOKE) \cite{Wieczorek2015}, which has been shown to be a robust method for studying the ultrafast magnetization dynamics of thin films, bulk materials, and bi- and multi-layer structures \zit{Traeger1992, Eschenlohr2016, Battiato2010, Stiehl2022}. 
It monitors the rotation of a linearly polarized light beam after reflection from the sample surface, which provides information about the spatially averaged magnetization dynamics in the material. 
By accessing the complex MOKE response, i.e.~the Kerr rotation and Kerr ellipticity, the individual magnetic responses of two different layers of selected multilayer structures can be disentangled. 
The resulting depth sensitive information \zit{Traeger1992, Eschenlohr2016,Wieczorek2015} provides insight into the spin current transport across interfaces and thus to the 
efficiency of optically-induced spin current injection into another (magnetic or non-magnetic) material. 
Unfortunately, the complex Kerr response does not contain enough information to follow the depth-dependent dynamics of ultrafast spin currents in a single magnetic material. Such information would be useful, for example, to study the correlation between spin-dependent particle and heat transport, to experimentally quantify ballistic, superdiffusive and diffusive spin transport \cite{Nenno2017, Jiang2022, Salvatella2016}, or to determine the spin current propagation length in a single material.
\par
In this paper, we introduce a new approach to studying ultrafast spin and heat transport in a single material by combining theoretical simulations and experimental method of \textit{angle-resolved Complex MOKE}. For the exemplary case of a homogeneous \SI{40}{\nano\meter} nickel film, we first predict the influence of ballistic and diffusive transport on the spatially dependent ultrafast magnetization dynamics of an inhomogeneous laser-heated Ni film using the thermodynamic $\mu$T model \zit{Ashok2022, Mueller2014}. These results are the basis for calculating the expected time-resolved complex MOKE response (Kerr rotation and ellipticity) for different angles of incidence of the probe beam. We find a clear variation in the angle-resolved Kerr-ellipticity traces in the case of diffusive transport, which is absent in the case of ballistic transport. Our predictions are compared with angle-resolved Complex MOKE data obtained for a similar sample, which indicate that the influence of ballistic transport on the ultrafast magnetization dynamics of a \SI{40}{\nano\meter} Ni film is not dominant.
\par
Section \ref{sec transport} studies the influence of diffusive and ballistic transport on spatially-resolved magnetization dynamics. Section \ref{sec cmoketheo} describes the relation between the CMOKE signal and the calculated spatially-resolved magnetization dynamics. Section \ref{sec theoresulcmoke} presents calculated probe-angle dependent Kerr-rotation and -ellipticity dynamics. Section \ref{sec exptsetup} describes the experimental setup and presents the measured CMOKE dynamics. Finally, section \ref{sec conclusion} concludes.
%\vspace{-0.6cm}
\section{Simulating the influence of Diffusive- and Ballistic transport in ultrafast magnetization dynamics}
\label{sec transport}

We start with the theoretical description of the depth-dependent magnetization dynamics $m(z, t)$ of an itinerant ferromagnet after laser excitation using the thermodynamic $\mu$T-model \cite{Mueller2014}. These simulations take into account spin-dependent heat transport, charge transport, Seebeck- and Peltier-effects \cite{Yamazaki2020, Seifert2018}. A full description of the model as well as all selected transport parameters are provided in our previous work in \zit{Ashok2022}. The optical excitation is described by an ultrashort Gaussian laser pulse with Full-width-at-half-max of $\SI{50}{\femto\second}$ and an absorbed fluence in Ni of $\SI{20}{m\joule{c\meter}^{-2}}$. The penetration depth of $\SI{15}{\nano\meter}$ in Ni is used to model the spatial absorption profile of the IR laser pulses ($\SI{800}{\nano\meter}$). 
\par
\begin{figure}[tbh]
    %\centering
    \begin{center}
     \hspace*{-.7cm}  
     \includegraphics[width=0.5\textwidth]{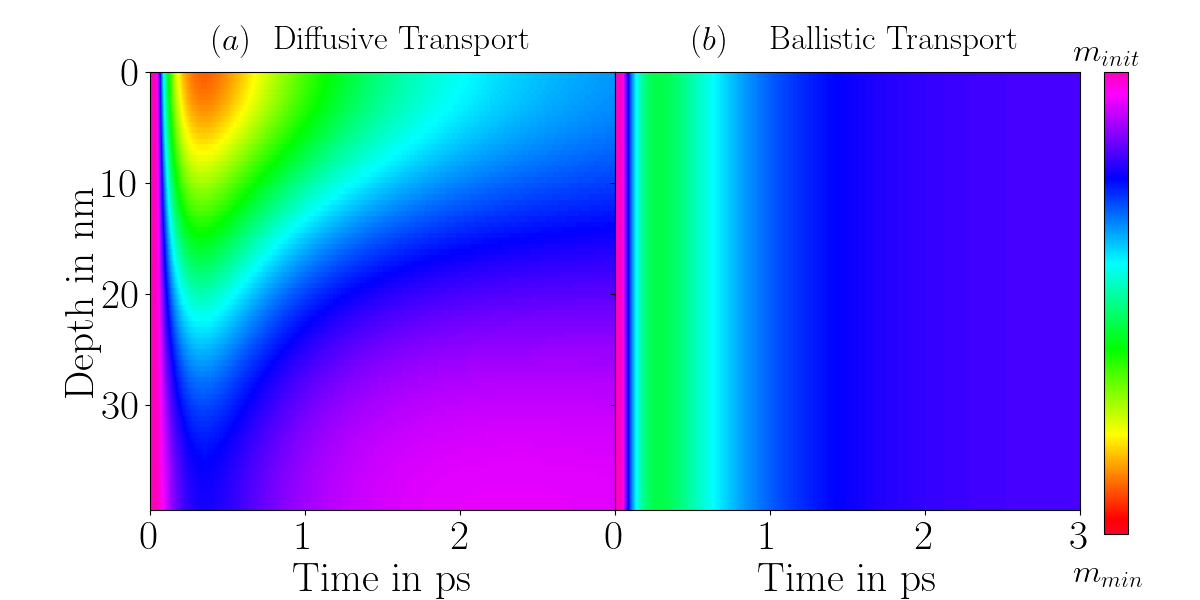}
    \end{center}
    %need to figure out how to insert large dpi figures into latex
    \caption{Theoretical results for magnetization dynamics in a $\SI{40}{\nano\meter}$ thick Nickel film in dependence on time and depth. The irradiated surface is at the top of the figure. The magnetization is color coded between $m_\text{init}$ and $m_\text{min}$, which denote the initial and minimal magnetization, respectively.  The cases of  simulated diffusive transport (a) and mimicked ballistic transport (b) are considered. In the former case, spatial gradients appear and persist within the material well into few picoseconds.  In the latter case, the assumed spatially homogeneous energy distribution leads to a homogeneous magnetization dynamics.}
    \label{fig:heatmap results}
\end{figure}
Fig.~\ref{fig:heatmap results}(a) shows the calculated magnetization dynamics $m(z, t)$ as a function of depth and time.  
The position of the irradiated surface corresponds to a depth of $0\,$nm and is located at the top of the plot. The color code indicates the relative magnitude of the magnetization between the initial magnetization $m_{init}$ and the extreme minimum $m_{min}$ that is reached within the film. 
The spatial depth profile of the optical excitation follows the Beer-Lambert's law and is responsible for the initial gradients in energy distributions. This inhomogeneity in the energy distribution leads to gradients in the magnetization dynamics that persist into the picosecond regime in the case of diffusive transport considered in the calculation underlying Fig.~\ref{fig:heatmap results}(a). 
\par
In contrast, ballistic transport leads to an almost instantaneous transformation of the initially created spatial gradients in the energy distribution deposited by the ultrashort laser pulse \cite{Hohlfeld2000}.  To mimic this situation, we consider an initially homogeneous energy distribution leading to a depth independent magnetization dynamics as illustrated in Fig.~\ref{fig:heatmap results}(b). 

\section{Simulating Kerr-response using spatially-resolved magnetization dynamics}
\label{sec cmoketheo}
Based on the depth-dependent magnetization dynamics $m(z,t)$, we now turn to the prediction of the time-dependent complex magneto-optical Kerr response of the magnetic thick film. In general, the complex Kerr response consists of two contributions, the Kerr rotation (real part) and the Kerr ellipticity (imaginary part). The Kerr rotation describes the rotation of the light polarization vector of a linearly polarized (probe) beam after its reflection from the magnetic film. On the other hand, the Kerr ellipticity corresponds to the degree of ellipticity of the probe beam resulting from the interaction of light with the magnetic material \zit{Kuch2015}.
\par
The complex Kerr response itself depends not only on the depth-dependent magnetization of the film, but also on the properties of the probe beam (angle of incidence, photon energy), the material (complex refractive index), and the light-matter interaction (magneto-optical constants) \cite{Schafer2007, Hamrle2002}. 
However, since the material parameters can be considered constant and time-independent in most cases, the transient changes in the complex Kerr response reflect the depth-dependent magnetization dynamics of magnetic films. 
\par
The simplest situation occurs for the Kerr response of ultrathin magnetic films \cite{Straub1996} with thicknesses less than the penetration depth of the pump and probe beams. 
In these cases, the pump beam leads to a homogeneous energy deposition inside the material and thus to depth-independent magnetization dynamics, while the reflected probe beam measures the averaged magnetization dynamics with an equal contribution from each $z$-position (depth) inside the material. 
Thus, the complex Kerr responsible for an ultrathin magnetic film can be treated as the Kerr response of the surface layer.
\par
\begin{figure}[t]
\hspace*{-0.8cm}
    %\centering
    \includegraphics[width=0.6\textwidth]{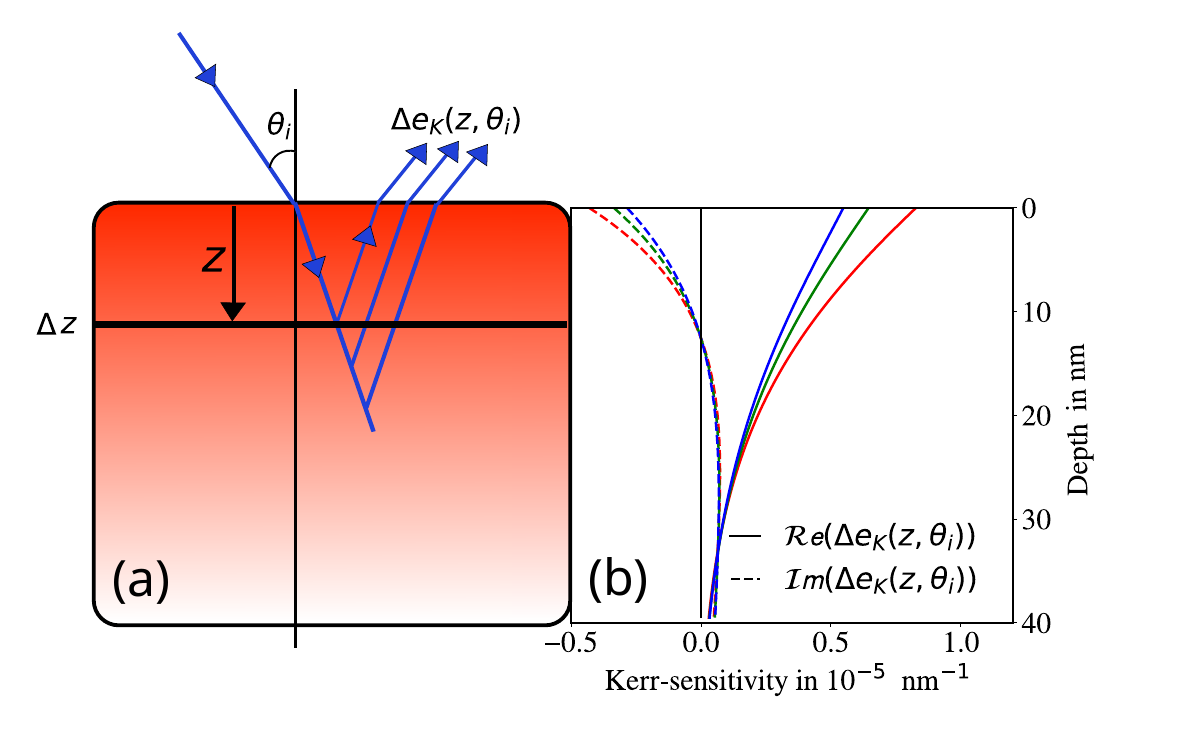}
    \caption{ The total complex Kerr-response of the magnetic film is calculated through an integration of depth-dependent differential contributions, see Eq.~(\ref{eq int kerr}). (a): Sketch of the penetration and virtual reflections at different depths. (b): Real (solid lines) and imaginary parts (dashed lines) of the Kerr-sensitivity function according to Eq.~(\ref{eq_kerrsensitivity}). Here exemplary curves for various probe-angles and a film thickness of $\SI{40}{\nano\meter}$ are shown. While the real-part monotonically decreases with depth, the imaginary part increases up to a certain depth and then monotonically decreases (see Supplementary materials).
    }
    \label{fig:effective_thin_layer}
\end{figure}

In contrast, the contributions of the magnetization dynamics in different depths inside the material must be explicitly considered as soon as the film thickness exceeds the penetration depth of the probe beam \cite{Traeger1992}. This can be done by dividing the magnetic thick film into infinitesimally thin slices with thickness $\Delta z$ as illustrated in Fig.~\ref{fig:effective_thin_layer}(a). The Kerr-response of such an infinitesimally thin slice in the material at a depth $z$ with respect to the surface can be described by the differential CMOKE-response 
\begin{equation}\label{eq diffkerr}
\Delta e^{\text{K}}_{\text{bulk}}(z,t, \theta_{i}) = \Delta e_{\text{K}}(z, \theta_{i})  \cdot m(z,t) \cdot \Delta z
\end{equation}
with a depth-dependent magnetization dynamics $m(z,t)$ and a {complex} Kerr-sensitivity function $\Delta e_{\text{K}}(z, \theta_{i})$. This function weights the contributions from the individual depth-dependent slices to the overall Kerr response of the material and can be calculated by \cite{Wenzel1997, Traeger1992, Kuch2015}
\begin{equation}\label{eq_kerrsensitivity}
\Delta e_{\text{K}}(z, \theta_{i}) = -K_{\text{bulk}}(\theta_{i})  \cdot 2 \eta  \cdot \phi (2z)\enspace,
\end{equation}
where $K_{\text{bulk}}(\theta_{i})$ is the complex Kerr coefficient for a uniformly magnetized thick film. It depends on the angle of incidence of the probe beam $\theta_{i}$, the Voigt-constant, and transmission coefficients for s- and p-polarized light \cite{Kuch2015, herrmann2006}. The phase factor 
\begin{equation}{
\phi (z) = e^{\eta z}~~~~\text{with  }\eta = k_0  \cdot i \cos  \theta_{i}  \cdot n_1 }
\end{equation}
takes into account the effect of depth $z$, the complex refractive index $n_1$ of the film and the wavenumber $k_{0} = 2\pi / \lambda$, with $\lambda$ being the wavelength of the probe beam.
\par
Exemplary depth-dependent real and imaginary parts of the complex Kerr sensitivity function $\Delta e_{\text{K}}(z)$ are shown in Fig~\ref{fig:effective_thin_layer}(b) for a $\SI{40}{\nano\meter}$ thick film (see also Fig.~\ref{fig:cmplx_kerr_sensitivity} in the Supplementary materials). 
The real part of the Kerr sensitivity function (solid lines) decreases monotonically with increasing distance from the surface. 
This monotonic profile is almost independent of the angle of incidence of the probe beam (probe-angles). 
In contrast, the imaginary part of the Kerr sensitivity function (dashed lines) shows a non-monotonous behavior with a maximum at a characteristic depth of about $20\,$nm. 
Crucially, we also find a point of intersection of the imaginary part of the Kerr sensitivity function for different probe angles. 
This indicates notable differences in the depth-dependent sensitivity profiles in the imaginary part of the Kerr response.

In order to determine the total complex Kerr-response $K^{\text{total}}$ of a magnetic films, we take advantage of the principle of linear superposition and calculate the sum of CMOKE responses from infinitesimally thin slices within the material \cite{Kuch2015} as
\begin{eqnarray}
{K^{\text{total}}}(t, \theta_{i}) &=& \int_{0}^{L} \Delta e_{\text{K}}(z, \theta_{i})\cdot m(z,t) \, dz \, \label{eq int kerr}
\end{eqnarray}
with $L$ denoting the total thickness of the film. Thus, by \textit{weighting} the spatially-resolved magnetization dynamics with depth-dependent Kerr-sensitivity function, we obtain the probe angle dependent Complex Kerr-response, equivalently, the probe-angle dependent Kerr-rotation and Kerr-ellipticity dynamics.
\par
We now evaluate the dynamics of the total complex Kerr response $K^{\text{total}}(\theta_{i}, t)$ for an arbitrary probe angle $\theta_{i}$ and spatially-resolved magnetization $m(z,t)$. The real part of the total complex Kerr response corresponds to Kerr rotation and the imaginary part corresponds to Kerr ellipticity. When the spatially-resolved magnetization is time dependent, Kerr rotation and Kerr ellipticity also evolve with time. These are termed Kerr-rotation dynamics and Kerr-ellipticity dynamics, respectively.

\section{Complex Kerr dynamics for homogeneous and inhomogeneous magnetization dynamics}
\label{sec theoresulcmoke}
The time and angle resolved Kerr response of a $40\,$nm Ni film can be predicted by calculating the complex Kerr response $K^{\text{total}}(\theta_{i}, t)$ using the spatially resolved magnetization dynamics shown in Fig.~\ref{fig:heatmap results}. For the longitudinal MOKE geometry \cite{Keay1968}, the material and probe angle specific constants $K_{\text{bulk}}$ are determined using the equation from \zit{Kuch2015},
\begin{equation}
    K_{\text{bulk}} = \frac{i Q_{V} \sin(\theta_{i}) }{4 \cos(\theta_{i}) \cos(\theta_{r})} t_{ss}^{01} t_{pp}^{01}. 
\end{equation}
Here, $\theta_{i} \text{ and } \theta_{r}$ are the angles of incidence and refraction, respectively. $t_{ss}^{01}$ and $t_{pp}^{01}$ are the transmission coefficients for the s- and p-polarized components of the probe, respectively. The complex number $Q_{V}$ is the Voigt constant whose value is taken from the literature \zit{herrmann2006}. All other simulation parameters are given in the table \ref{table: sim_vals} in the Supplementary materials.
\par
The Fig.~\ref{fig:theo results}(a) shows the Kerr rotation dynamics (top figure) and the Kerr ellipticity dynamics (bottom figure) for the case of diffusive transport. The angle of incidence of the probe beam is varied between ${30^{\circ}}$ and ${50^{\circ}}$. The initial values of Kerr rotation and ellipticity before optical excitation are normalized to $1$. The normalized Kerr rotation curves are (almost) identical for all probe angles. This observation can be attributed to the rapid reduction of the real part of the Kerr sensitivity function with increasing distance from the surface, as discussed above. As a result, the Kerr rotation dynamics is insensitive to the gradients in the magnetization.
\par
In contrast, we find a clear qualitative difference in the angle dependence of the time-dependent Kerr-ellipticity traces, which are shown in the bottom figure of the left panel in Fig.~\ref{fig:theo results}. For selected probe angles, the Kerr ellipticity increases monotonically after optical excitation before decreasing again, while the time evolution of the Kerr ellipticity is completely opposite for other probe angles. This strong probe angle dependence is due to the imaginary part of the Kerr-sensitivity function, which shows characteristic differences in its depth profile for different probe angles. In particular, its shape enhances the magnetization dynamics for large distances from the surface relative to that of the surface region (see Fig.~\ref{fig:cumulative_signal_diffusive} in the Supplementary materials).
\par
\onecolumngrid
\vspace*{0.5cm}
\begin{figure}[!ht]
    %\centering
    \includegraphics[width=0.45\textwidth]{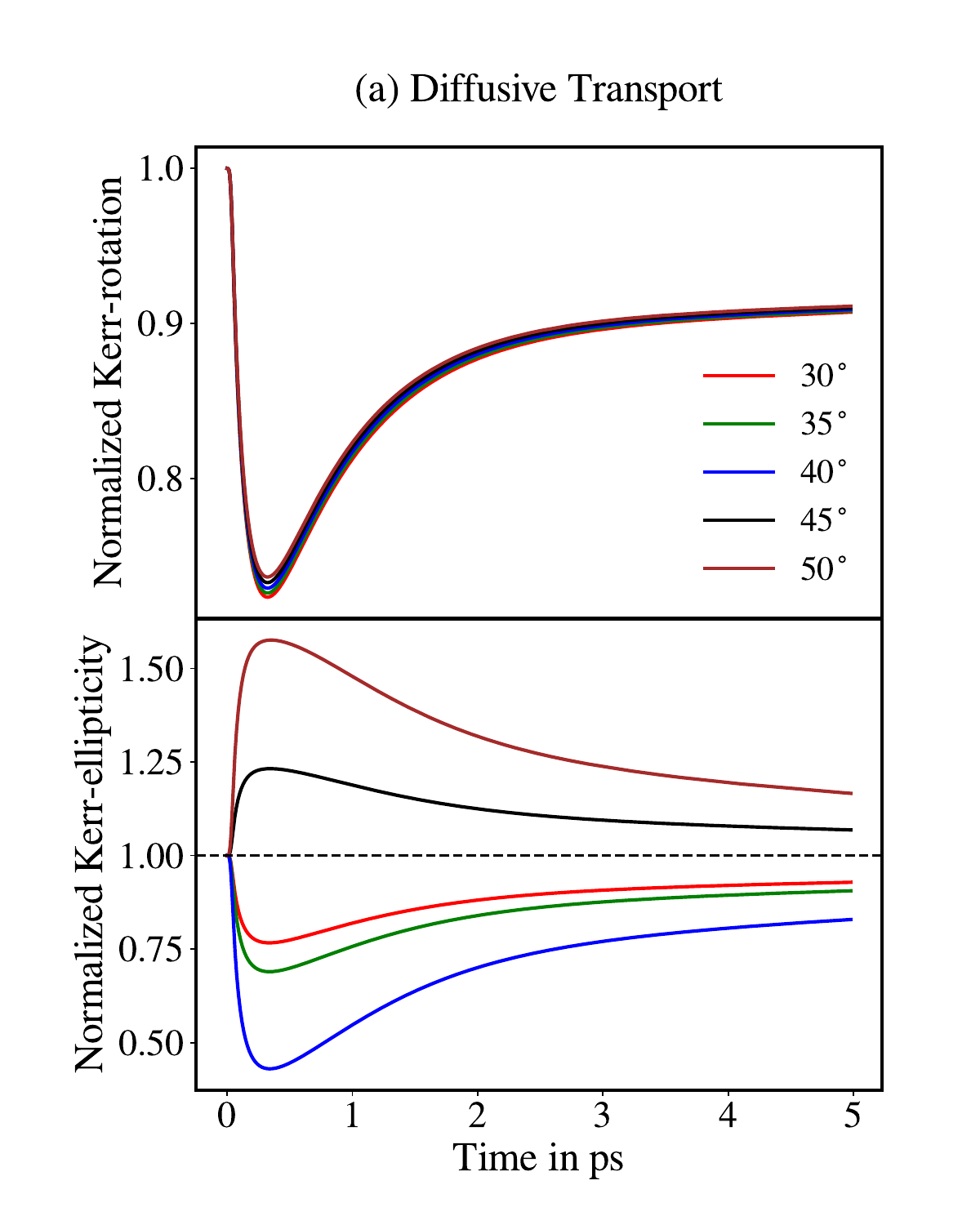}
    \includegraphics[width=0.45\textwidth]{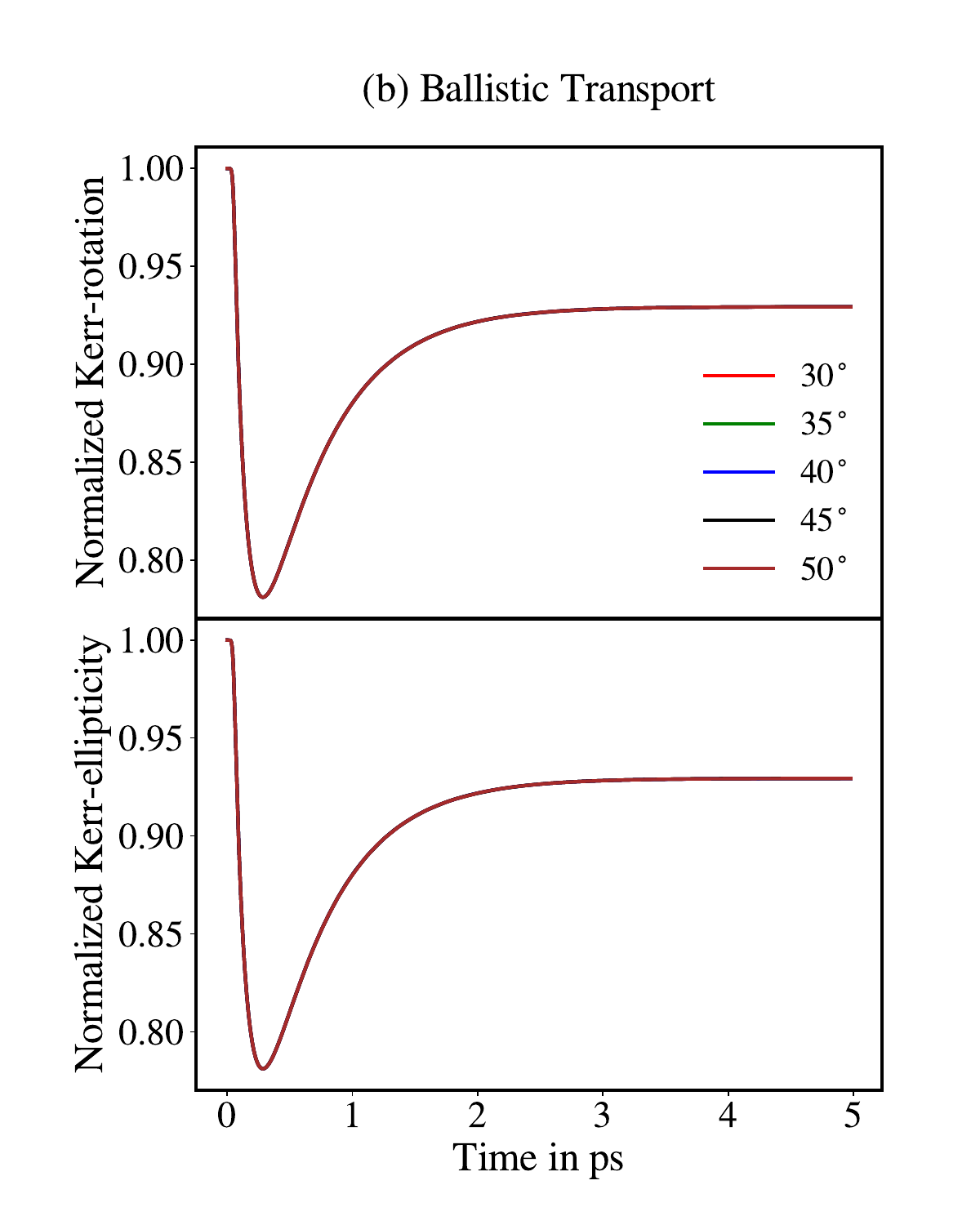}
    \caption{ Simulations of Kerr-rotation and -ellipticity dynamics as calculated for various probe angles for a Nickel film after ultrafast laser excitation. (a): Realistic pump penetration depth as well as diffusive transport have been considered. The magnetization is spatially inhomogeneous during the time of calculation. (b):
    Homogeneous excitation. Pump penetration has been set to a large value while the transport coefficients have the same magnitude as that of the diffusive transport case. This is to mimic ballistic transport leading to equilibration of gradients.
    While the Kerr-rotation curves (top figures) show only negligible differences between the two transport cases, the Kerr-ellipticity (bottom figures) reveals a strong sensitivity on gradients in magnetization, hence to transport mechanisms.  
    }
    \label{fig:theo results}
\end{figure}
%\vspace{1cm}
\twocolumngrid
For comparison, we show the complex Kerr response of the thick Ni film for the case of ballistic transport in Fig.~\ref{fig:theo results}(b). The probe angles are again varied between ${30^{\circ}}$ and ${50^{\circ}}$. The absence of spatial inhomogeneities in the magnetization leads to an absence of probe angle dependence of the Kerr rotation and ellipticity dynamics.
\par
Overall, our calculations of the time-resolved complex Kerr response for inhomogeneous (diffusive transport) and homogeneous (ballistic transport) magnetization dynamics in a magnetic material allow us to draw two conclusions: First, the Kerr rotation shows only marginal sensitivity to the depth-dependent magnetization dynamics and is thus insensitive to transport-induced gradients in magnetization. Second, the probe-angle dependent Kerr-ellipticity dynamics provides a signature that allows us to distinguish between ballistic and diffusive spin transport within a single magnetic material.
\par
In the following, we will confront our predictions with probe-angle dependent CMOKE experiments conducted for a \SI{40}{\nano\meter} nickel film.
\par	
	
\vspace*{1cm}
\twocolumngrid
\section{Angle resolved Complex Kerr Experiments on a Ni thick film}
\label{sec exptsetup}

\begin{figure}[ht]
        \centering
        \includegraphics[width=0.4\textwidth]{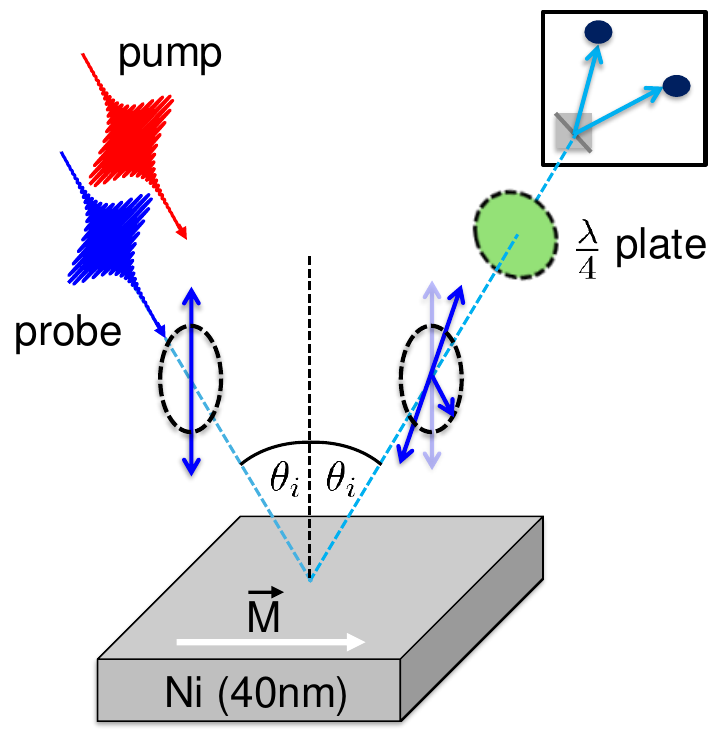}
        \caption{Pump-Probe setup with quarter waveplate}
        \label{fig:experimental_setup}
    \end{figure}

The time- and angle-resolved complex Kerr experiments are performed in a longitudinal Kerr geometry using s-polarised radiation for both the pump and probe beams. The corresponding experimental geometry is shown in Fig.~\ref{fig:experimental_setup}. The $40\,$nm Ni sample grown on an insulating MgO substrate is excited by an infrared pump pulse at a wavelength of $800\,$nm ($1.55\,$eV, pulse duration $<50\,$fs). The depth-dependent changes in magnetisation were monitored by changing the light polarisation (rotation and ellipticity) of the initially linearly polarised probe beam at a wavelength of $400\,$nm ($3.1\,$eV, pulse duration $<50\,$fs). The spot size of the probe beam on the sample surfaces was less than $200\,\mu$m, which is significantly smaller than the spot size of the pump beam. This configuration allows us to exclude any influence of lateral gradients in the deposited energy on the recorded magnetisation dynamics. 

The most important parameter of our collinear pump-probe experiment (the angle between the pump and probe beams is $\approx 1^{\circ}$) is the angle of incidence $\theta_i$ with respect to the sample surface normal. We vary the angle of incidence of both beams between $30^\circ$ and $50^\circ$ in steps of $5^\circ$. For each angle of incidence, we record time-dependent traces for Kerr rotation and Kerr ellipticity using a balanced photodetector in combination with a quarter-wave plate. Together these traces represent the time- and angle-resolved complex Kerr response of the Ni thick film.
\par
\begin{figure}[tb]
    \centering
    \includegraphics[width=0.5\textwidth]{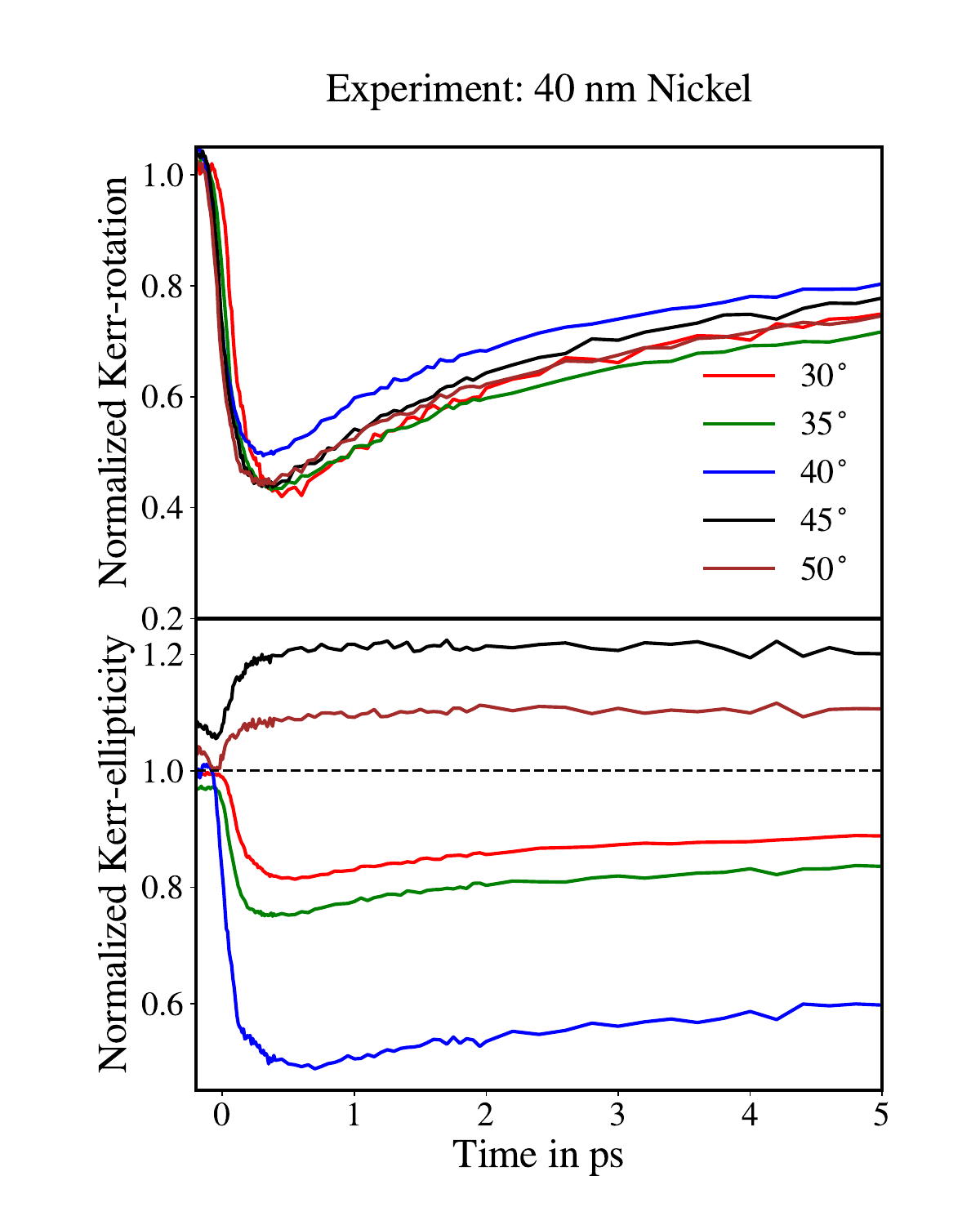}
    \caption{Experimental results for Kerr-rotation and -ellipticity dynamics measured for various probe angles.}
    \label{fig:results exp}
\end{figure}

The experimental results of our time- and angle-resolved complex Kerr study for a $40\,$nm Ni film are shown in Fig.~\ref{fig:results exp}. All traces were obtained with an almost identical applied laser fluence. % of F$=XX\,\frac{mJ}{cm^2}$
In addition, the effective penetration depth of the pump beam is almost identical for all angles of incidence considered in our study, as can be estimated from Snells law of refraction at the interface~\cite{Werner2009}. This leads to an almost identical depth-dependent excitation profile for all time-dependent complex Kerr traces recorded in our experiment.

We start with the Kerr rotation traces for different angles of incidence, shown in the upper part of Fig.~\ref{fig:results exp}. All traces have been normalised to the Kerr rotation signal before optical excitation. Within our experimental accuracy, all time-dependent Kerr rotation traces show exactly the same temporal evolution, independent of the angle of incidence. This observation confirms our theoretical predictions that the Kerr rotation is quite robust to small variations in the experimental geometry, i.e. the angle of incidence between the incoming pump and probe beams and the sample surfaces.
\par
In contrast, we find strong and qualitative differences in the temporal evolution of the Kerr ellipticity for different angles of incidence, as shown in the lower part of Fig.~\ref{fig:results exp}. Again, the Kerr ellipticity signal before optical excitation has been normalised to unity. For certain angles of incidence, the Kerr ellipticity signal after optical excitation increases, while for other angles of incidence it decreases. Such variations in the angle-dependent Kerr ellipticity traces were also observed in our model simulations and were attributed to depth-dependent gradients in the magnetisation dynamics. In particular, the depth-dependent inhomogeneities in the magnetization exist as long as the Kerr ellipticity traces merge and follow the same temporal evolution.
\par
For the exemplary case of the $40\,$nm Ni film, our experimental results thus indicate the existence of strong gradients in the depth-dependent magnetization dynamics after optical excitation, which persist at least for $5\,$ps. These long-lasting depth-dependent magnetization dynamics are a strong indication that diffusive transport is the dominant transport mechanism in this magnetic thick film. In particular, the contribution of ballistic transport is not strong enough to lead to homogeneous magnetization dynamics within the Ni thick film.

\section{Conclusion}
\label{sec conclusion}

In conclusion, our work has demonstrated the potential of time- and angle-resolved complex MOKE to identify the evolution of depth-dependent gradients and inhomogeneities in the magnetization dynamics of chemically and structurally homogeneous magnetic thick films after optical excitation using fs light pulses. Our main results are based on the theoretical and experimental investigation of the temporal evolution of Kerr rotation and Kerr ellipticity for different incident angles of the probe beam. We show that the time- and angle-resolved Kerr rotation traces do not show any angle-dependency, neither for the case of depth-dependent nor the depth-independent magnetisation dynamics, making the Kerr rotation a highly robust experimental observable. On the other hand, the Kerr ellipticity shows a clear angular dependence in the case of spatial gradients and diffusive transport within the magnetic materials, which is absent in the case of spatially homogeneous magnetization dynamics and ballistic transport. These insights allowed us to identify diffusive transport as the dominant mode of transport within an optically excited $40\,$nm Ni film. In this way, our work provides a new avenue for experimentally exploring transport phenomena in homogeneous magnetic materials on ultrafast timescales. 

\section{Acknowledgements}

\section{Funding}

The research leading to this study was funded by the Deutsche Forschungsgemeinschaft (DFG, German Research Foundation)—TRR 173—268565370 Spin+X (Project No. B03).

\twocolumngrid

\bibliography{refs}

%apsrev4-2.bst 2019-01-14 (MD) hand-edited version of apsrev4-1.bst
%Control: key (0)
%Control: author (8) initials jnrlst
%Control: editor formatted (1) identically to author
%Control: production of article title (0) allowed
%Control: page (0) single
%Control: year (1) truncated
%Control: production of eprint (0) enabled
\begin{thebibliography}{42}%
\makeatletter
\providecommand \@ifxundefined [1]{%
 \@ifx{#1\undefined}
}%
\providecommand \@ifnum [1]{%
 \ifnum #1\expandafter \@firstoftwo
 \else \expandafter \@secondoftwo
 \fi
}%
\providecommand \@ifx [1]{%
 \ifx #1\expandafter \@firstoftwo
 \else \expandafter \@secondoftwo
 \fi
}%
\providecommand \natexlab [1]{#1}%
\providecommand \enquote  [1]{``#1''}%
\providecommand \bibnamefont  [1]{#1}%
\providecommand \bibfnamefont [1]{#1}%
\providecommand \citenamefont [1]{#1}%
\providecommand \href@noop [0]{\@secondoftwo}%
\providecommand \href [0]{\begingroup \@sanitize@url \@href}%
\providecommand \@href[1]{\@@startlink{#1}\@@href}%
\providecommand \@@href[1]{\endgroup#1\@@endlink}%
\providecommand \@sanitize@url [0]{\catcode `\\12\catcode `\$12\catcode `\&12\catcode `\#12\catcode `\^12\catcode `\_12\catcode `\%12\relax}%
\providecommand \@@startlink[1]{}%
\providecommand \@@endlink[0]{}%
\providecommand \url  [0]{\begingroup\@sanitize@url \@url }%
\providecommand \@url [1]{\endgroup\@href {#1}{\urlprefix }}%
\providecommand \urlprefix  [0]{URL }%
\providecommand \Eprint [0]{\href }%
\providecommand \doibase [0]{https://doi.org/}%
\providecommand \selectlanguage [0]{\@gobble}%
\providecommand \bibinfo  [0]{\@secondoftwo}%
\providecommand \bibfield  [0]{\@secondoftwo}%
\providecommand \translation [1]{[#1]}%
\providecommand \BibitemOpen [0]{}%
\providecommand \bibitemStop [0]{}%
\providecommand \bibitemNoStop [0]{.\EOS\space}%
\providecommand \EOS [0]{\spacefactor3000\relax}%
\providecommand \BibitemShut  [1]{\csname bibitem#1\endcsname}%
\let\auto@bib@innerbib\@empty
%</preamble>
\bibitem [{\citenamefont {Kent}\ and\ \citenamefont {Worledge}(2015)}]{Kent2015}%
  \BibitemOpen
  \bibfield  {author} {\bibinfo {author} {\bibfnamefont {A.~D.}\ \bibnamefont {Kent}}\ and\ \bibinfo {author} {\bibfnamefont {D.~C.}\ \bibnamefont {Worledge}},\ }\bibfield  {title} {\bibinfo {title} {A new spin on magnetic memories},\ }\href@noop {} {\bibfield  {journal} {\bibinfo  {journal} {Nat. Nanotechnol.}\ }\textbf {\bibinfo {volume} {10}},\ \bibinfo {pages} {187} (\bibinfo {year} {2015})}\BibitemShut {NoStop}%
\bibitem [{\citenamefont {van~'t Erve}\ \emph {et~al.}(2009)\citenamefont {van~'t Erve}, \citenamefont {Awo-Affouda}, \citenamefont {Hanbicki}, \citenamefont {Li}, \citenamefont {Thompson},\ and\ \citenamefont {Jonker}}]{Erve2009}%
  \BibitemOpen
  \bibfield  {author} {\bibinfo {author} {\bibfnamefont {O.~M.~J.}\ \bibnamefont {van~'t Erve}}, \bibinfo {author} {\bibfnamefont {C.}~\bibnamefont {Awo-Affouda}}, \bibinfo {author} {\bibfnamefont {A.~T.}\ \bibnamefont {Hanbicki}}, \bibinfo {author} {\bibfnamefont {C.~H.}\ \bibnamefont {Li}}, \bibinfo {author} {\bibfnamefont {P.~E.}\ \bibnamefont {Thompson}},\ and\ \bibinfo {author} {\bibfnamefont {B.~T.}\ \bibnamefont {Jonker}},\ }\bibfield  {title} {\bibinfo {title} {Information processing with pure spin currents in silicon: Spin injection, extraction, manipulation, and detection},\ }\href {https://doi.org/10.1109/TED.2009.2027975} {\bibfield  {journal} {\bibinfo  {journal} {IEEE Trans.~Electron Devices}\ }\textbf {\bibinfo {volume} {56}},\ \bibinfo {pages} {2343} (\bibinfo {year} {2009})}\BibitemShut {NoStop}%
\bibitem [{\citenamefont {Kampfrath}\ \emph {et~al.}(2013)\citenamefont {Kampfrath}, \citenamefont {Battiato}, \citenamefont {Maldonado}, \citenamefont {Eilers}, \citenamefont {N{\"o}tzold}, \citenamefont {M{\"a}hrlein}, \citenamefont {Zbarsky}, \citenamefont {Freimuth}, \citenamefont {Mokrousov}, \citenamefont {Bl{\"u}gel}, \citenamefont {Wolf}, \citenamefont {Radu}, \citenamefont {Oppeneer},\ and\ \citenamefont {M{\"u}nzenberg}}]{Kampfrath2013}%
  \BibitemOpen
  \bibfield  {author} {\bibinfo {author} {\bibfnamefont {T.}~\bibnamefont {Kampfrath}}, \bibinfo {author} {\bibfnamefont {M.}~\bibnamefont {Battiato}}, \bibinfo {author} {\bibfnamefont {P.}~\bibnamefont {Maldonado}}, \bibinfo {author} {\bibfnamefont {G.}~\bibnamefont {Eilers}}, \bibinfo {author} {\bibfnamefont {J.}~\bibnamefont {N{\"o}tzold}}, \bibinfo {author} {\bibfnamefont {S.}~\bibnamefont {M{\"a}hrlein}}, \bibinfo {author} {\bibfnamefont {V.}~\bibnamefont {Zbarsky}}, \bibinfo {author} {\bibfnamefont {F.}~\bibnamefont {Freimuth}}, \bibinfo {author} {\bibfnamefont {Y.}~\bibnamefont {Mokrousov}}, \bibinfo {author} {\bibfnamefont {S.}~\bibnamefont {Bl{\"u}gel}}, \bibinfo {author} {\bibfnamefont {M.}~\bibnamefont {Wolf}}, \bibinfo {author} {\bibfnamefont {I.}~\bibnamefont {Radu}}, \bibinfo {author} {\bibfnamefont {P.~M.}\ \bibnamefont {Oppeneer}},\ and\ \bibinfo {author} {\bibfnamefont {M.}~\bibnamefont {M{\"u}nzenberg}},\ }\bibfield  {title} {\bibinfo {title} {Terahertz spin current pulses controlled by
  magnetic heterostructures},\ }\href@noop {} {\bibfield  {journal} {\bibinfo  {journal} {Nat. Nanotechnol.}\ }\textbf {\bibinfo {volume} {8}},\ \bibinfo {pages} {256} (\bibinfo {year} {2013})}\BibitemShut {NoStop}%
\bibitem [{\citenamefont {Hoffmann}\ and\ \citenamefont {Bader}(2015)}]{Hoffmann2015}%
  \BibitemOpen
  \bibfield  {author} {\bibinfo {author} {\bibfnamefont {A.}~\bibnamefont {Hoffmann}}\ and\ \bibinfo {author} {\bibfnamefont {S.~D.}\ \bibnamefont {Bader}},\ }\bibfield  {title} {\bibinfo {title} {Opportunities at the frontiers of spintronics},\ }\href@noop {} {\bibfield  {journal} {\bibinfo  {journal} {Phys. Rev. Appl.}\ }\textbf {\bibinfo {volume} {4}},\ \bibinfo {pages} {047001} (\bibinfo {year} {2015})}\BibitemShut {NoStop}%
\bibitem [{\citenamefont {Chappert}\ \emph {et~al.}(2007)\citenamefont {Chappert}, \citenamefont {Fert},\ and\ \citenamefont {Van~Dau}}]{Chappert2007}%
  \BibitemOpen
  \bibfield  {author} {\bibinfo {author} {\bibfnamefont {C.}~\bibnamefont {Chappert}}, \bibinfo {author} {\bibfnamefont {A.}~\bibnamefont {Fert}},\ and\ \bibinfo {author} {\bibfnamefont {F.~N.}\ \bibnamefont {Van~Dau}},\ }\bibfield  {title} {\bibinfo {title} {The emergence of spin electronics in data storage},\ }\href@noop {} {\bibfield  {journal} {\bibinfo  {journal} {Nature materials}\ }\textbf {\bibinfo {volume} {6}},\ \bibinfo {pages} {813} (\bibinfo {year} {2007})}\BibitemShut {NoStop}%
\bibitem [{\citenamefont {Lebrun}\ \emph {et~al.}(2018)\citenamefont {Lebrun}, \citenamefont {Ross}, \citenamefont {Bender}, \citenamefont {Qaiumzadeh}, \citenamefont {Baldrati}, \citenamefont {Cramer}, \citenamefont {Brataas}, \citenamefont {Duine},\ and\ \citenamefont {Kl{\"a}ui}}]{Lebrun2018}%
  \BibitemOpen
  \bibfield  {author} {\bibinfo {author} {\bibfnamefont {R.}~\bibnamefont {Lebrun}}, \bibinfo {author} {\bibfnamefont {A.}~\bibnamefont {Ross}}, \bibinfo {author} {\bibfnamefont {S.}~\bibnamefont {Bender}}, \bibinfo {author} {\bibfnamefont {A.}~\bibnamefont {Qaiumzadeh}}, \bibinfo {author} {\bibfnamefont {L.}~\bibnamefont {Baldrati}}, \bibinfo {author} {\bibfnamefont {J.}~\bibnamefont {Cramer}}, \bibinfo {author} {\bibfnamefont {A.}~\bibnamefont {Brataas}}, \bibinfo {author} {\bibfnamefont {R.}~\bibnamefont {Duine}},\ and\ \bibinfo {author} {\bibfnamefont {M.}~\bibnamefont {Kl{\"a}ui}},\ }\bibfield  {title} {\bibinfo {title} {Tunable long-distance spin transport in a crystalline antiferromagnetic iron oxide},\ }\href@noop {} {\bibfield  {journal} {\bibinfo  {journal} {Nature}\ }\textbf {\bibinfo {volume} {561}},\ \bibinfo {pages} {222} (\bibinfo {year} {2018})}\BibitemShut {NoStop}%
\bibitem [{\citenamefont {Pfeiffer}\ \emph {et~al.}(2021)\citenamefont {Pfeiffer}, \citenamefont {Reeve}, \citenamefont {Elphick}, \citenamefont {Hirohata},\ and\ \citenamefont {Kl{\"a}ui}}]{Pfeiffer2021}%
  \BibitemOpen
  \bibfield  {author} {\bibinfo {author} {\bibfnamefont {A.}~\bibnamefont {Pfeiffer}}, \bibinfo {author} {\bibfnamefont {R.~M.}\ \bibnamefont {Reeve}}, \bibinfo {author} {\bibfnamefont {K.}~\bibnamefont {Elphick}}, \bibinfo {author} {\bibfnamefont {A.}~\bibnamefont {Hirohata}},\ and\ \bibinfo {author} {\bibfnamefont {M.}~\bibnamefont {Kl{\"a}ui}},\ }\bibfield  {title} {\bibinfo {title} {Revealing the importance of interfaces for pure spin current transport},\ }\href@noop {} {\bibfield  {journal} {\bibinfo  {journal} {Phys. rev. res.}\ }\textbf {\bibinfo {volume} {3}},\ \bibinfo {pages} {023110} (\bibinfo {year} {2021})}\BibitemShut {NoStop}%
\bibitem [{\citenamefont {Reimers}\ \emph {et~al.}(2023)\citenamefont {Reimers}, \citenamefont {Lytvynenko}, \citenamefont {Niu}, \citenamefont {Golias}, \citenamefont {Sarpi}, \citenamefont {Veiga}, \citenamefont {Denneulin}, \citenamefont {Kovacs}, \citenamefont {Dunin-Borkowski}, \citenamefont {Bl{\"a}{\ss}er}, \citenamefont {Kl{\"a}ui},\ and\ \citenamefont {Jourdan}}]{Reimers2023}%
  \BibitemOpen
  \bibfield  {author} {\bibinfo {author} {\bibfnamefont {S.}~\bibnamefont {Reimers}}, \bibinfo {author} {\bibfnamefont {Y.}~\bibnamefont {Lytvynenko}}, \bibinfo {author} {\bibfnamefont {Y.}~\bibnamefont {Niu}}, \bibinfo {author} {\bibfnamefont {E.}~\bibnamefont {Golias}}, \bibinfo {author} {\bibfnamefont {B.}~\bibnamefont {Sarpi}}, \bibinfo {author} {\bibfnamefont {L.}~\bibnamefont {Veiga}}, \bibinfo {author} {\bibfnamefont {T.}~\bibnamefont {Denneulin}}, \bibinfo {author} {\bibfnamefont {A.}~\bibnamefont {Kovacs}}, \bibinfo {author} {\bibfnamefont {R.~E.}\ \bibnamefont {Dunin-Borkowski}}, \bibinfo {author} {\bibfnamefont {J.}~\bibnamefont {Bl{\"a}{\ss}er}}, \bibinfo {author} {\bibfnamefont {M.}~\bibnamefont {Kl{\"a}ui}},\ and\ \bibinfo {author} {\bibfnamefont {M.}~\bibnamefont {Jourdan}},\ }\bibfield  {title} {\bibinfo {title} {Current-driven writing process in antiferromagnetic mn2au for memory applications},\ }\href@noop {} {\bibfield  {journal} {\bibinfo  {journal} {Nat. Commun.}\ }\textbf {\bibinfo
  {volume} {14}},\ \bibinfo {pages} {1861} (\bibinfo {year} {2023})}\BibitemShut {NoStop}%
\bibitem [{\citenamefont {Cheng}\ \emph {et~al.}(2016)\citenamefont {Cheng}, \citenamefont {Xiao},\ and\ \citenamefont {Brataas}}]{Cheng2016}%
  \BibitemOpen
  \bibfield  {author} {\bibinfo {author} {\bibfnamefont {R.}~\bibnamefont {Cheng}}, \bibinfo {author} {\bibfnamefont {D.}~\bibnamefont {Xiao}},\ and\ \bibinfo {author} {\bibfnamefont {A.}~\bibnamefont {Brataas}},\ }\bibfield  {title} {\bibinfo {title} {Terahertz antiferromagnetic spin hall nano oscillator},\ }\href@noop {} {\bibfield  {journal} {\bibinfo  {journal} {Phys. Rev. Lett.}\ }\textbf {\bibinfo {volume} {116}},\ \bibinfo {pages} {207603} (\bibinfo {year} {2016})}\BibitemShut {NoStop}%
\bibitem [{\citenamefont {Rudolf}\ \emph {et~al.}(2012)\citenamefont {Rudolf}, \citenamefont {La-O-Vorakiat}, \citenamefont {Battiato}, \citenamefont {Adam}, \citenamefont {Shaw}, \citenamefont {Turgut}, \citenamefont {Maldonado}, \citenamefont {Mathias}, \citenamefont {Grychtol}, \citenamefont {Nembach}, \citenamefont {Silva}, \citenamefont {Aeschlimann}, \citenamefont {Kapteyn}, \citenamefont {Murnane~M.}, \citenamefont {Schneider},\ and\ \citenamefont {Oppeneer}}]{Rudolf2012}%
  \BibitemOpen
  \bibfield  {author} {\bibinfo {author} {\bibfnamefont {D.}~\bibnamefont {Rudolf}}, \bibinfo {author} {\bibfnamefont {C.}~\bibnamefont {La-O-Vorakiat}}, \bibinfo {author} {\bibfnamefont {M.}~\bibnamefont {Battiato}}, \bibinfo {author} {\bibfnamefont {R.}~\bibnamefont {Adam}}, \bibinfo {author} {\bibfnamefont {J.~M.}\ \bibnamefont {Shaw}}, \bibinfo {author} {\bibfnamefont {E.}~\bibnamefont {Turgut}}, \bibinfo {author} {\bibfnamefont {P.}~\bibnamefont {Maldonado}}, \bibinfo {author} {\bibfnamefont {S.}~\bibnamefont {Mathias}}, \bibinfo {author} {\bibfnamefont {P.}~\bibnamefont {Grychtol}}, \bibinfo {author} {\bibfnamefont {H.~T.}\ \bibnamefont {Nembach}}, \bibinfo {author} {\bibfnamefont {T.~J.}\ \bibnamefont {Silva}}, \bibinfo {author} {\bibfnamefont {M.}~\bibnamefont {Aeschlimann}}, \bibinfo {author} {\bibfnamefont {H.~C.}\ \bibnamefont {Kapteyn}}, \bibinfo {author} {\bibfnamefont {M.}~\bibnamefont {Murnane~M.}}, \bibinfo {author} {\bibfnamefont {C.~M.}\ \bibnamefont {Schneider}},\ and\ \bibinfo {author}
  {\bibfnamefont {P.~M.}\ \bibnamefont {Oppeneer}},\ }\bibfield  {title} {\bibinfo {title} {Ultrafast magnetization enhancement in metallic multilayers driven by superdiffusive spin current},\ }\href@noop {} {\bibfield  {journal} {\bibinfo  {journal} {Nat. Commun.}\ }\textbf {\bibinfo {volume} {3}},\ \bibinfo {pages} {1037} (\bibinfo {year} {2012})}\BibitemShut {NoStop}%
\bibitem [{\citenamefont {Melnikov}\ \emph {et~al.}(2022)\citenamefont {Melnikov}, \citenamefont {Brandt}, \citenamefont {Liebing}, \citenamefont {Ribow}, \citenamefont {Mertig},\ and\ \citenamefont {Woltersdorf}}]{Melnikov2022}%
  \BibitemOpen
  \bibfield  {author} {\bibinfo {author} {\bibfnamefont {A.}~\bibnamefont {Melnikov}}, \bibinfo {author} {\bibfnamefont {L.}~\bibnamefont {Brandt}}, \bibinfo {author} {\bibfnamefont {N.}~\bibnamefont {Liebing}}, \bibinfo {author} {\bibfnamefont {M.}~\bibnamefont {Ribow}}, \bibinfo {author} {\bibfnamefont {I.}~\bibnamefont {Mertig}},\ and\ \bibinfo {author} {\bibfnamefont {G.}~\bibnamefont {Woltersdorf}},\ }\bibfield  {title} {\bibinfo {title} {Ultrafast spin transport and control of spin current pulse shape in metallic multilayers},\ }\href@noop {} {\bibfield  {journal} {\bibinfo  {journal} {Phys. Rev. B}\ }\textbf {\bibinfo {volume} {106}},\ \bibinfo {pages} {104417} (\bibinfo {year} {2022})}\BibitemShut {NoStop}%
\bibitem [{\citenamefont {Choi}\ \emph {et~al.}(2014)\citenamefont {Choi}, \citenamefont {Min}, \citenamefont {Lee},\ and\ \citenamefont {Cahill}}]{Choi2014}%
  \BibitemOpen
  \bibfield  {author} {\bibinfo {author} {\bibfnamefont {G.-M.}\ \bibnamefont {Choi}}, \bibinfo {author} {\bibfnamefont {B.-C.}\ \bibnamefont {Min}}, \bibinfo {author} {\bibfnamefont {K.-J.}\ \bibnamefont {Lee}},\ and\ \bibinfo {author} {\bibfnamefont {D.~G.}\ \bibnamefont {Cahill}},\ }\bibfield  {title} {\bibinfo {title} {Spin current generated by thermally driven ultrafast demagnetization},\ }\href@noop {} {\bibfield  {journal} {\bibinfo  {journal} {Nat. Commun.}\ }\textbf {\bibinfo {volume} {5}},\ \bibinfo {pages} {4334} (\bibinfo {year} {2014})}\BibitemShut {NoStop}%
\bibitem [{\citenamefont {Fert{\'e}}\ \emph {et~al.}(2019)\citenamefont {Fert{\'e}}, \citenamefont {Bergeard}, \citenamefont {Malinowski}, \citenamefont {Terrier}, \citenamefont {Le~Guyader}, \citenamefont {Holldack}, \citenamefont {Hehn},\ and\ \citenamefont {Boeglin}}]{Ferte2019}%
  \BibitemOpen
  \bibfield  {author} {\bibinfo {author} {\bibfnamefont {T.}~\bibnamefont {Fert{\'e}}}, \bibinfo {author} {\bibfnamefont {N.}~\bibnamefont {Bergeard}}, \bibinfo {author} {\bibfnamefont {G.}~\bibnamefont {Malinowski}}, \bibinfo {author} {\bibfnamefont {E.}~\bibnamefont {Terrier}}, \bibinfo {author} {\bibfnamefont {L.}~\bibnamefont {Le~Guyader}}, \bibinfo {author} {\bibfnamefont {K.}~\bibnamefont {Holldack}}, \bibinfo {author} {\bibfnamefont {M.}~\bibnamefont {Hehn}},\ and\ \bibinfo {author} {\bibfnamefont {C.}~\bibnamefont {Boeglin}},\ }\bibfield  {title} {\bibinfo {title} {Ultrafast demagnetization in buried co80dy20 as fingerprint of hot-electron transport},\ }\href@noop {} {\bibfield  {journal} {\bibinfo  {journal} {J. Magn. Magn. Mater.}\ }\textbf {\bibinfo {volume} {485}},\ \bibinfo {pages} {320} (\bibinfo {year} {2019})}\BibitemShut {NoStop}%
\bibitem [{\citenamefont {Beaurepaire}\ \emph {et~al.}(1996)\citenamefont {Beaurepaire}, \citenamefont {Merle}, \citenamefont {Daunois},\ and\ \citenamefont {Bigot}}]{Beaurepaire1996}%
  \BibitemOpen
  \bibfield  {author} {\bibinfo {author} {\bibfnamefont {E.}~\bibnamefont {Beaurepaire}}, \bibinfo {author} {\bibfnamefont {J.-C.}\ \bibnamefont {Merle}}, \bibinfo {author} {\bibfnamefont {A.}~\bibnamefont {Daunois}},\ and\ \bibinfo {author} {\bibfnamefont {J.-Y.}\ \bibnamefont {Bigot}},\ }\bibfield  {title} {\bibinfo {title} {Ultrafast spin dynamics in ferromagnetic nickel},\ }\href@noop {} {\bibfield  {journal} {\bibinfo  {journal} {Phys. Rev. Lett.}\ }\textbf {\bibinfo {volume} {76}},\ \bibinfo {pages} {4250} (\bibinfo {year} {1996})}\BibitemShut {NoStop}%
\bibitem [{\citenamefont {Dutta}\ \emph {et~al.}(2022)\citenamefont {Dutta}, \citenamefont {Narayan~Panda}, \citenamefont {Sinha}, \citenamefont {Choudhury},\ and\ \citenamefont {Barman}}]{Dutta2022}%
  \BibitemOpen
  \bibfield  {author} {\bibinfo {author} {\bibfnamefont {S.}~\bibnamefont {Dutta}}, \bibinfo {author} {\bibfnamefont {S.}~\bibnamefont {Narayan~Panda}}, \bibinfo {author} {\bibfnamefont {J.}~\bibnamefont {Sinha}}, \bibinfo {author} {\bibfnamefont {S.}~\bibnamefont {Choudhury}},\ and\ \bibinfo {author} {\bibfnamefont {A.}~\bibnamefont {Barman}},\ }\bibfield  {title} {\bibinfo {title} {Role of spin transport through the $\beta$-ta/co20fe60b20 interface on its ultrafast demagnetization: Implications for ultra-high-speed spin-orbitronic devices},\ }\href@noop {} {\bibfield  {journal} {\bibinfo  {journal} {ACS Appl. Nano Mater.}\ } (\bibinfo {year} {2022})}\BibitemShut {NoStop}%
\bibitem [{\citenamefont {Melnikov}\ \emph {et~al.}(2011)\citenamefont {Melnikov}, \citenamefont {Razdolski}, \citenamefont {Wehling}, \citenamefont {Papaioannou}, \citenamefont {Roddatis}, \citenamefont {Fumagalli}, \citenamefont {Aktsipetrov}, \citenamefont {Lichtenstein},\ and\ \citenamefont {Bovensiepen}}]{Melnikov2011}%
  \BibitemOpen
  \bibfield  {author} {\bibinfo {author} {\bibfnamefont {A.}~\bibnamefont {Melnikov}}, \bibinfo {author} {\bibfnamefont {I.}~\bibnamefont {Razdolski}}, \bibinfo {author} {\bibfnamefont {T.~O.}\ \bibnamefont {Wehling}}, \bibinfo {author} {\bibfnamefont {E.~T.}\ \bibnamefont {Papaioannou}}, \bibinfo {author} {\bibfnamefont {V.}~\bibnamefont {Roddatis}}, \bibinfo {author} {\bibfnamefont {P.}~\bibnamefont {Fumagalli}}, \bibinfo {author} {\bibfnamefont {O.}~\bibnamefont {Aktsipetrov}}, \bibinfo {author} {\bibfnamefont {A.~I.}\ \bibnamefont {Lichtenstein}},\ and\ \bibinfo {author} {\bibfnamefont {U.}~\bibnamefont {Bovensiepen}},\ }\bibfield  {title} {\bibinfo {title} {Ultrafast transport of laser-excited spin-polarized carriers in au/fe/mgo (001)},\ }\href@noop {} {\bibfield  {journal} {\bibinfo  {journal} {Phys. Rev. Lett.}\ }\textbf {\bibinfo {volume} {107}},\ \bibinfo {pages} {076601} (\bibinfo {year} {2011})}\BibitemShut {NoStop}%
\bibitem [{\citenamefont {Schellekens}\ \emph {et~al.}(2014)\citenamefont {Schellekens}, \citenamefont {Kuiper}, \citenamefont {De~Wit},\ and\ \citenamefont {Koopmans}}]{Schellekens2014}%
  \BibitemOpen
  \bibfield  {author} {\bibinfo {author} {\bibfnamefont {A.}~\bibnamefont {Schellekens}}, \bibinfo {author} {\bibfnamefont {K.}~\bibnamefont {Kuiper}}, \bibinfo {author} {\bibfnamefont {R.}~\bibnamefont {De~Wit}},\ and\ \bibinfo {author} {\bibfnamefont {B.}~\bibnamefont {Koopmans}},\ }\bibfield  {title} {\bibinfo {title} {Ultrafast spin-transfer torque driven by femtosecond pulsed-laser excitation},\ }\href@noop {} {\bibfield  {journal} {\bibinfo  {journal} {Nat. Commun.}\ }\textbf {\bibinfo {volume} {5}},\ \bibinfo {pages} {4333} (\bibinfo {year} {2014})}\BibitemShut {NoStop}%
\bibitem [{\citenamefont {Chirac}\ \emph {et~al.}(2020)\citenamefont {Chirac}, \citenamefont {Chauleau}, \citenamefont {Thibaudeau}, \citenamefont {Gomonay},\ and\ \citenamefont {Viret}}]{Chirac2020}%
  \BibitemOpen
  \bibfield  {author} {\bibinfo {author} {\bibfnamefont {T.}~\bibnamefont {Chirac}}, \bibinfo {author} {\bibfnamefont {J.-Y.}\ \bibnamefont {Chauleau}}, \bibinfo {author} {\bibfnamefont {P.}~\bibnamefont {Thibaudeau}}, \bibinfo {author} {\bibfnamefont {O.}~\bibnamefont {Gomonay}},\ and\ \bibinfo {author} {\bibfnamefont {M.}~\bibnamefont {Viret}},\ }\bibfield  {title} {\bibinfo {title} {Ultrafast antiferromagnetic switching in nio induced by spin transfer torques},\ }\href@noop {} {\bibfield  {journal} {\bibinfo  {journal} {Phys. Rev. B}\ }\textbf {\bibinfo {volume} {102}},\ \bibinfo {pages} {134415} (\bibinfo {year} {2020})}\BibitemShut {NoStop}%
\bibitem [{\citenamefont {Gerrits}\ \emph {et~al.}(2002)\citenamefont {Gerrits}, \citenamefont {Van Den~Berg}, \citenamefont {Hohlfeld}, \citenamefont {B{\"a}r},\ and\ \citenamefont {Rasing}}]{Gerrits2002}%
  \BibitemOpen
  \bibfield  {author} {\bibinfo {author} {\bibfnamefont {T.}~\bibnamefont {Gerrits}}, \bibinfo {author} {\bibfnamefont {H.}~\bibnamefont {Van Den~Berg}}, \bibinfo {author} {\bibfnamefont {J.}~\bibnamefont {Hohlfeld}}, \bibinfo {author} {\bibfnamefont {L.}~\bibnamefont {B{\"a}r}},\ and\ \bibinfo {author} {\bibfnamefont {T.}~\bibnamefont {Rasing}},\ }\bibfield  {title} {\bibinfo {title} {Ultrafast precessional magnetization reversal by picosecond magnetic field pulse shaping},\ }\href@noop {} {\bibfield  {journal} {\bibinfo  {journal} {Nature}\ }\textbf {\bibinfo {volume} {418}},\ \bibinfo {pages} {509} (\bibinfo {year} {2002})}\BibitemShut {NoStop}%
\bibitem [{\citenamefont {Remy}\ \emph {et~al.}(2023)\citenamefont {Remy}, \citenamefont {Hohlfeld}, \citenamefont {Verg{\`e}s}, \citenamefont {Le~Guen}, \citenamefont {Gorchon}, \citenamefont {Malinowski}, \citenamefont {Mangin},\ and\ \citenamefont {Hehn}}]{Remy2023}%
  \BibitemOpen
  \bibfield  {author} {\bibinfo {author} {\bibfnamefont {Q.}~\bibnamefont {Remy}}, \bibinfo {author} {\bibfnamefont {J.}~\bibnamefont {Hohlfeld}}, \bibinfo {author} {\bibfnamefont {M.}~\bibnamefont {Verg{\`e}s}}, \bibinfo {author} {\bibfnamefont {Y.}~\bibnamefont {Le~Guen}}, \bibinfo {author} {\bibfnamefont {J.}~\bibnamefont {Gorchon}}, \bibinfo {author} {\bibfnamefont {G.}~\bibnamefont {Malinowski}}, \bibinfo {author} {\bibfnamefont {S.}~\bibnamefont {Mangin}},\ and\ \bibinfo {author} {\bibfnamefont {M.}~\bibnamefont {Hehn}},\ }\bibfield  {title} {\bibinfo {title} {Accelerating ultrafast magnetization reversal by non-local spin transfer},\ }\href@noop {} {\bibfield  {journal} {\bibinfo  {journal} {Nat. Commun.}\ }\textbf {\bibinfo {volume} {14}},\ \bibinfo {pages} {445} (\bibinfo {year} {2023})}\BibitemShut {NoStop}%
\bibitem [{\citenamefont {Krieger}\ \emph {et~al.}(2017)\citenamefont {Krieger}, \citenamefont {Elliott}, \citenamefont {M{\"u}ller}, \citenamefont {Singh}, \citenamefont {Dewhurst}, \citenamefont {Gross},\ and\ \citenamefont {Sharma}}]{Krieger2017}%
  \BibitemOpen
  \bibfield  {author} {\bibinfo {author} {\bibfnamefont {K.}~\bibnamefont {Krieger}}, \bibinfo {author} {\bibfnamefont {P.}~\bibnamefont {Elliott}}, \bibinfo {author} {\bibfnamefont {T.}~\bibnamefont {M{\"u}ller}}, \bibinfo {author} {\bibfnamefont {N.}~\bibnamefont {Singh}}, \bibinfo {author} {\bibfnamefont {J.}~\bibnamefont {Dewhurst}}, \bibinfo {author} {\bibfnamefont {E.}~\bibnamefont {Gross}},\ and\ \bibinfo {author} {\bibfnamefont {S.}~\bibnamefont {Sharma}},\ }\bibfield  {title} {\bibinfo {title} {Ultrafast demagnetization in bulk versus thin films: an ab initio study},\ }\href@noop {} {\bibfield  {journal} {\bibinfo  {journal} {J. Phys. Condens. Matter}\ }\textbf {\bibinfo {volume} {29}},\ \bibinfo {pages} {224001} (\bibinfo {year} {2017})}\BibitemShut {NoStop}%
\bibitem [{\citenamefont {Wieczorek}\ \emph {et~al.}(2015)\citenamefont {Wieczorek}, \citenamefont {Eschenlohr}, \citenamefont {Weidtmann}, \citenamefont {R{\"o}sner}, \citenamefont {Bergeard}, \citenamefont {Tarasevitch}, \citenamefont {Wehling},\ and\ \citenamefont {Bovensiepen}}]{Wieczorek2015}%
  \BibitemOpen
  \bibfield  {author} {\bibinfo {author} {\bibfnamefont {J.}~\bibnamefont {Wieczorek}}, \bibinfo {author} {\bibfnamefont {A.}~\bibnamefont {Eschenlohr}}, \bibinfo {author} {\bibfnamefont {B.}~\bibnamefont {Weidtmann}}, \bibinfo {author} {\bibfnamefont {M.}~\bibnamefont {R{\"o}sner}}, \bibinfo {author} {\bibfnamefont {N.}~\bibnamefont {Bergeard}}, \bibinfo {author} {\bibfnamefont {A.}~\bibnamefont {Tarasevitch}}, \bibinfo {author} {\bibfnamefont {T.}~\bibnamefont {Wehling}},\ and\ \bibinfo {author} {\bibfnamefont {U.}~\bibnamefont {Bovensiepen}},\ }\bibfield  {title} {\bibinfo {title} {Separation of ultrafast spin currents and spin-flip scattering in co/cu (001) driven by femtosecond laser excitation employing the complex magneto-optical kerr effect},\ }\href@noop {} {\bibfield  {journal} {\bibinfo  {journal} {Phys. Rev. B}\ }\textbf {\bibinfo {volume} {92}},\ \bibinfo {pages} {174410} (\bibinfo {year} {2015})}\BibitemShut {NoStop}%
\bibitem [{\citenamefont {Traeger}\ \emph {et~al.}(1992)\citenamefont {Traeger}, \citenamefont {Wenzel},\ and\ \citenamefont {Hubert}}]{Traeger1992}%
  \BibitemOpen
  \bibfield  {author} {\bibinfo {author} {\bibfnamefont {G.}~\bibnamefont {Traeger}}, \bibinfo {author} {\bibfnamefont {L.}~\bibnamefont {Wenzel}},\ and\ \bibinfo {author} {\bibfnamefont {A.}~\bibnamefont {Hubert}},\ }\bibfield  {title} {\bibinfo {title} {Computer experiments on the information depth and the figure of merit in magnetooptics},\ }\href@noop {} {\bibfield  {journal} {\bibinfo  {journal} {Phys. Status Solidi (a)}\ }\textbf {\bibinfo {volume} {131}},\ \bibinfo {pages} {201} (\bibinfo {year} {1992})}\BibitemShut {NoStop}%
\bibitem [{\citenamefont {Eschenlohr}\ \emph {et~al.}(2016)\citenamefont {Eschenlohr}, \citenamefont {Wieczorek}, \citenamefont {Chen}, \citenamefont {Weidtmann}, \citenamefont {R{\"o}sner}, \citenamefont {Bergeard}, \citenamefont {Tarasevitch}, \citenamefont {Wehling},\ and\ \citenamefont {Bovensiepen}}]{Eschenlohr2016}%
  \BibitemOpen
  \bibfield  {author} {\bibinfo {author} {\bibfnamefont {A.}~\bibnamefont {Eschenlohr}}, \bibinfo {author} {\bibfnamefont {J.}~\bibnamefont {Wieczorek}}, \bibinfo {author} {\bibfnamefont {J.}~\bibnamefont {Chen}}, \bibinfo {author} {\bibfnamefont {B.}~\bibnamefont {Weidtmann}}, \bibinfo {author} {\bibfnamefont {M.}~\bibnamefont {R{\"o}sner}}, \bibinfo {author} {\bibfnamefont {N.}~\bibnamefont {Bergeard}}, \bibinfo {author} {\bibfnamefont {A.}~\bibnamefont {Tarasevitch}}, \bibinfo {author} {\bibfnamefont {T.~O.}\ \bibnamefont {Wehling}},\ and\ \bibinfo {author} {\bibfnamefont {U.}~\bibnamefont {Bovensiepen}},\ }\bibfield  {title} {\bibinfo {title} {Analyzing ultrafast laser-induced demagnetization in co/cu (001) via the depth sensitivity of the time-resolved transversal magneto-optical kerr effect}\ }(\bibinfo {organization} {Ultrafast Phenomena and Nanophotonics XX, SPIE},\ \bibinfo {year} {2016})\ pp.\ \bibinfo {pages} {154--160}\BibitemShut {NoStop}%
\bibitem [{\citenamefont {Battiato}\ \emph {et~al.}(2010)\citenamefont {Battiato}, \citenamefont {Carva},\ and\ \citenamefont {Oppeneer}}]{Battiato2010}%
  \BibitemOpen
  \bibfield  {author} {\bibinfo {author} {\bibfnamefont {M.}~\bibnamefont {Battiato}}, \bibinfo {author} {\bibfnamefont {K.}~\bibnamefont {Carva}},\ and\ \bibinfo {author} {\bibfnamefont {P.~M.}\ \bibnamefont {Oppeneer}},\ }\bibfield  {title} {\bibinfo {title} {Superdiffusive spin transport as a mechanism of ultrafast demagnetization},\ }\href@noop {} {\bibfield  {journal} {\bibinfo  {journal} {Phys. Rev. Lett.}\ }\textbf {\bibinfo {volume} {105}},\ \bibinfo {pages} {027203} (\bibinfo {year} {2010})}\BibitemShut {NoStop}%
\bibitem [{\citenamefont {Stiehl}\ \emph {et~al.}(2022)\citenamefont {Stiehl}, \citenamefont {Weber}, \citenamefont {Seibel}, \citenamefont {Hoefer}, \citenamefont {Weber}, \citenamefont {Nenno}, \citenamefont {Schneider}, \citenamefont {Rethfeld}, \citenamefont {Stadtm{\"u}ller},\ and\ \citenamefont {Aeschlimann}}]{Stiehl2022}%
  \BibitemOpen
  \bibfield  {author} {\bibinfo {author} {\bibfnamefont {M.}~\bibnamefont {Stiehl}}, \bibinfo {author} {\bibfnamefont {M.}~\bibnamefont {Weber}}, \bibinfo {author} {\bibfnamefont {C.}~\bibnamefont {Seibel}}, \bibinfo {author} {\bibfnamefont {J.}~\bibnamefont {Hoefer}}, \bibinfo {author} {\bibfnamefont {S.~T.}\ \bibnamefont {Weber}}, \bibinfo {author} {\bibfnamefont {D.~M.}\ \bibnamefont {Nenno}}, \bibinfo {author} {\bibfnamefont {H.~C.}\ \bibnamefont {Schneider}}, \bibinfo {author} {\bibfnamefont {B.}~\bibnamefont {Rethfeld}}, \bibinfo {author} {\bibfnamefont {B.}~\bibnamefont {Stadtm{\"u}ller}},\ and\ \bibinfo {author} {\bibfnamefont {M.}~\bibnamefont {Aeschlimann}},\ }\bibfield  {title} {\bibinfo {title} {Role of primary and secondary processes in the ultrafast spin dynamics of nickel},\ }\href@noop {} {\bibfield  {journal} {\bibinfo  {journal} {Appl. Phys. Lett.}\ }\textbf {\bibinfo {volume} {120}},\ \bibinfo {pages} {062410} (\bibinfo {year} {2022})}\BibitemShut {NoStop}%
\bibitem [{\citenamefont {Nenno}\ \emph {et~al.}(2017)\citenamefont {Nenno}, \citenamefont {Weber},\ and\ \citenamefont {Schneider}}]{Nenno2017}%
  \BibitemOpen
  \bibfield  {author} {\bibinfo {author} {\bibfnamefont {D.}~\bibnamefont {Nenno}}, \bibinfo {author} {\bibfnamefont {M.}~\bibnamefont {Weber}},\ and\ \bibinfo {author} {\bibfnamefont {H.~C.}\ \bibnamefont {Schneider}},\ }\bibfield  {title} {\bibinfo {title} {Simulation of ultrafast spin-dependent hot-electron transport in metallic multilayers},\ }\href@noop {} {\bibfield  {journal} {\bibinfo  {journal} {Bulletin of the American Physical Society}\ }\textbf {\bibinfo {volume} {62}} (\bibinfo {year} {2017})}\BibitemShut {NoStop}%
\bibitem [{\citenamefont {Jiang}\ \emph {et~al.}(2022)\citenamefont {Jiang}, \citenamefont {Zhao}, \citenamefont {Chen}, \citenamefont {You}, \citenamefont {Lai},\ and\ \citenamefont {Zhao}}]{Jiang2022}%
  \BibitemOpen
  \bibfield  {author} {\bibinfo {author} {\bibfnamefont {T.}~\bibnamefont {Jiang}}, \bibinfo {author} {\bibfnamefont {X.}~\bibnamefont {Zhao}}, \bibinfo {author} {\bibfnamefont {Z.}~\bibnamefont {Chen}}, \bibinfo {author} {\bibfnamefont {Y.}~\bibnamefont {You}}, \bibinfo {author} {\bibfnamefont {T.}~\bibnamefont {Lai}},\ and\ \bibinfo {author} {\bibfnamefont {J.}~\bibnamefont {Zhao}},\ }\bibfield  {title} {\bibinfo {title} {Ultrafast dynamics of demagnetization in femn/mnga bilayer nanofilm structures via phonon transport},\ }\href@noop {} {\bibfield  {journal} {\bibinfo  {journal} {Nanomaterials}\ }\textbf {\bibinfo {volume} {12}},\ \bibinfo {pages} {4088} (\bibinfo {year} {2022})}\BibitemShut {NoStop}%
\bibitem [{\citenamefont {Salvatella}\ \emph {et~al.}(2016)\citenamefont {Salvatella}, \citenamefont {Gort}, \citenamefont {B{\"u}hlmann}, \citenamefont {D{\"a}ster}, \citenamefont {Vaterlaus},\ and\ \citenamefont {Acremann}}]{Salvatella2016}%
  \BibitemOpen
  \bibfield  {author} {\bibinfo {author} {\bibfnamefont {G.}~\bibnamefont {Salvatella}}, \bibinfo {author} {\bibfnamefont {R.}~\bibnamefont {Gort}}, \bibinfo {author} {\bibfnamefont {K.}~\bibnamefont {B{\"u}hlmann}}, \bibinfo {author} {\bibfnamefont {S.}~\bibnamefont {D{\"a}ster}}, \bibinfo {author} {\bibfnamefont {A.}~\bibnamefont {Vaterlaus}},\ and\ \bibinfo {author} {\bibfnamefont {Y.}~\bibnamefont {Acremann}},\ }\bibfield  {title} {\bibinfo {title} {Ultrafast demagnetization by hot electrons: Diffusion or super-diffusion?},\ }\href@noop {} {\bibfield  {journal} {\bibinfo  {journal} {Struct. Dyn.}\ }\textbf {\bibinfo {volume} {3}},\ \bibinfo {pages} {055101} (\bibinfo {year} {2016})}\BibitemShut {NoStop}%
\bibitem [{\citenamefont {Ashok}\ \emph {et~al.}(2022)\citenamefont {Ashok}, \citenamefont {Seibel}, \citenamefont {Weber}, \citenamefont {Briones},\ and\ \citenamefont {Rethfeld}}]{Ashok2022}%
  \BibitemOpen
  \bibfield  {author} {\bibinfo {author} {\bibfnamefont {S.}~\bibnamefont {Ashok}}, \bibinfo {author} {\bibfnamefont {C.}~\bibnamefont {Seibel}}, \bibinfo {author} {\bibfnamefont {S.~T.}\ \bibnamefont {Weber}}, \bibinfo {author} {\bibfnamefont {J.}~\bibnamefont {Briones}},\ and\ \bibinfo {author} {\bibfnamefont {B.}~\bibnamefont {Rethfeld}},\ }\bibfield  {title} {\bibinfo {title} {Influence of diffusive transport on ultrafast magnetization dynamics},\ }\href@noop {} {\bibfield  {journal} {\bibinfo  {journal} {Appl. Phys. Lett.}\ }\textbf {\bibinfo {volume} {120}},\ \bibinfo {pages} {142402} (\bibinfo {year} {2022})}\BibitemShut {NoStop}%
\bibitem [{\citenamefont {Mueller}\ and\ \citenamefont {Rethfeld}(2014)}]{Mueller2014}%
  \BibitemOpen
  \bibfield  {author} {\bibinfo {author} {\bibfnamefont {B.}~\bibnamefont {Mueller}}\ and\ \bibinfo {author} {\bibfnamefont {B.}~\bibnamefont {Rethfeld}},\ }\bibfield  {title} {\bibinfo {title} {Thermodynamic $\mu$t model of ultrafast magnetization dynamics},\ }\href@noop {} {\bibfield  {journal} {\bibinfo  {journal} {Phys. Rev. B}\ }\textbf {\bibinfo {volume} {90}},\ \bibinfo {pages} {144420} (\bibinfo {year} {2014})}\BibitemShut {NoStop}%
\bibitem [{\citenamefont {Yamazaki}\ \emph {et~al.}(2020)\citenamefont {Yamazaki}, \citenamefont {Iguchi}, \citenamefont {Ohkubo}, \citenamefont {Nagano},\ and\ \citenamefont {Uchida}}]{Yamazaki2020}%
  \BibitemOpen
  \bibfield  {author} {\bibinfo {author} {\bibfnamefont {T.}~\bibnamefont {Yamazaki}}, \bibinfo {author} {\bibfnamefont {R.}~\bibnamefont {Iguchi}}, \bibinfo {author} {\bibfnamefont {T.}~\bibnamefont {Ohkubo}}, \bibinfo {author} {\bibfnamefont {H.}~\bibnamefont {Nagano}},\ and\ \bibinfo {author} {\bibfnamefont {K.-i.}\ \bibnamefont {Uchida}},\ }\bibfield  {title} {\bibinfo {title} {Transient response of the spin peltier effect revealed by lock-in thermoreflectance measurements},\ }\href@noop {} {\bibfield  {journal} {\bibinfo  {journal} {Phys. Rev. B}\ }\textbf {\bibinfo {volume} {101}},\ \bibinfo {pages} {020415} (\bibinfo {year} {2020})}\BibitemShut {NoStop}%
\bibitem [{\citenamefont {Seifert}\ \emph {et~al.}(2018)\citenamefont {Seifert}, \citenamefont {Jaiswal}, \citenamefont {Barker}, \citenamefont {Weber}, \citenamefont {Razdolski}, \citenamefont {Cramer}, \citenamefont {Gueckstock}, \citenamefont {Maehrlein}, \citenamefont {Nadvornik}, \citenamefont {Watanabe}, \citenamefont {Ciccarelli}, \citenamefont {Melnikov}, \citenamefont {Gerhard}, \citenamefont {Muenzenberg}, \citenamefont {Goennenwein}, \citenamefont {Woltersdorf}, \citenamefont {Rethfeld}, \citenamefont {Brouwer}, \citenamefont {Wolf}, \citenamefont {Kl{\"a}ui},\ and\ \citenamefont {Kampfrath}}]{Seifert2018}%
  \BibitemOpen
  \bibfield  {author} {\bibinfo {author} {\bibfnamefont {T.~S.}\ \bibnamefont {Seifert}}, \bibinfo {author} {\bibfnamefont {S.}~\bibnamefont {Jaiswal}}, \bibinfo {author} {\bibfnamefont {J.}~\bibnamefont {Barker}}, \bibinfo {author} {\bibfnamefont {S.~T.}\ \bibnamefont {Weber}}, \bibinfo {author} {\bibfnamefont {I.}~\bibnamefont {Razdolski}}, \bibinfo {author} {\bibfnamefont {J.}~\bibnamefont {Cramer}}, \bibinfo {author} {\bibfnamefont {O.}~\bibnamefont {Gueckstock}}, \bibinfo {author} {\bibfnamefont {S.~F.}\ \bibnamefont {Maehrlein}}, \bibinfo {author} {\bibfnamefont {L.}~\bibnamefont {Nadvornik}}, \bibinfo {author} {\bibfnamefont {S.}~\bibnamefont {Watanabe}}, \bibinfo {author} {\bibfnamefont {C.}~\bibnamefont {Ciccarelli}}, \bibinfo {author} {\bibfnamefont {A.}~\bibnamefont {Melnikov}}, \bibinfo {author} {\bibfnamefont {J.}~\bibnamefont {Gerhard}}, \bibinfo {author} {\bibfnamefont {M.}~\bibnamefont {Muenzenberg}}, \bibinfo {author} {\bibfnamefont {S.~T.}\ \bibnamefont {Goennenwein}}, \bibinfo {author}
  {\bibfnamefont {G.}~\bibnamefont {Woltersdorf}}, \bibinfo {author} {\bibfnamefont {B.}~\bibnamefont {Rethfeld}}, \bibinfo {author} {\bibfnamefont {P.~W.}\ \bibnamefont {Brouwer}}, \bibinfo {author} {\bibfnamefont {M.}~\bibnamefont {Wolf}}, \bibinfo {author} {\bibfnamefont {M.}~\bibnamefont {Kl{\"a}ui}},\ and\ \bibinfo {author} {\bibfnamefont {T.}~\bibnamefont {Kampfrath}},\ }\bibfield  {title} {\bibinfo {title} {Femtosecond formation dynamics of the spin seebeck effect revealed by terahertz spectroscopy},\ }\href@noop {} {\bibfield  {journal} {\bibinfo  {journal} {Nat. Commun.}\ }\textbf {\bibinfo {volume} {9}},\ \bibinfo {pages} {2899} (\bibinfo {year} {2018})}\BibitemShut {NoStop}%
\bibitem [{\citenamefont {Hohlfeld}\ \emph {et~al.}(2000)\citenamefont {Hohlfeld}, \citenamefont {Wellershoff}, \citenamefont {G{\"u}dde}, \citenamefont {Conrad}, \citenamefont {J{\"a}hnke},\ and\ \citenamefont {Matthias}}]{Hohlfeld2000}%
  \BibitemOpen
  \bibfield  {author} {\bibinfo {author} {\bibfnamefont {J.}~\bibnamefont {Hohlfeld}}, \bibinfo {author} {\bibfnamefont {S.-S.}\ \bibnamefont {Wellershoff}}, \bibinfo {author} {\bibfnamefont {J.}~\bibnamefont {G{\"u}dde}}, \bibinfo {author} {\bibfnamefont {U.}~\bibnamefont {Conrad}}, \bibinfo {author} {\bibfnamefont {V.}~\bibnamefont {J{\"a}hnke}},\ and\ \bibinfo {author} {\bibfnamefont {E.}~\bibnamefont {Matthias}},\ }\bibfield  {title} {\bibinfo {title} {Electron and lattice dynamics following optical excitation of metals},\ }\href@noop {} {\bibfield  {journal} {\bibinfo  {journal} {Chemical physics}\ }\textbf {\bibinfo {volume} {251}},\ \bibinfo {pages} {237} (\bibinfo {year} {2000})}\BibitemShut {NoStop}%
\bibitem [{\citenamefont {Kuch}\ \emph {et~al.}(2015)\citenamefont {Kuch}, \citenamefont {Sch{\"a}fer}, \citenamefont {Fischer},\ and\ \citenamefont {Hillebrecht}}]{Kuch2015}%
  \BibitemOpen
  \bibfield  {author} {\bibinfo {author} {\bibfnamefont {W.}~\bibnamefont {Kuch}}, \bibinfo {author} {\bibfnamefont {R.}~\bibnamefont {Sch{\"a}fer}}, \bibinfo {author} {\bibfnamefont {P.}~\bibnamefont {Fischer}},\ and\ \bibinfo {author} {\bibfnamefont {F.~U.}\ \bibnamefont {Hillebrecht}},\ }\bibfield  {title} {\bibinfo {title} {Depth-sensitive conventional magneto-optical microscopy}\ }(\bibinfo  {publisher} {Book title: Magnetic Microscopy of Layered Structures, Springer},\ \bibinfo {year} {2015})\ pp.\ \bibinfo {pages} {97--140}\BibitemShut {NoStop}%
\bibitem [{\citenamefont {Sch{\"a}fer}(2007)}]{Schafer2007}%
  \BibitemOpen
  \bibfield  {author} {\bibinfo {author} {\bibfnamefont {R.}~\bibnamefont {Sch{\"a}fer}},\ }\bibfield  {title} {\bibinfo {title} {Investigation of domains and dynamics of domain walls by the magneto-optical k err-effect},\ }\href@noop {} {\bibfield  {journal} {\bibinfo  {journal} {Handbook of magnetism and advanced magnetic materials}\ } (\bibinfo {year} {2007})}\BibitemShut {NoStop}%
\bibitem [{\citenamefont {Hamrle}\ \emph {et~al.}(2002)\citenamefont {Hamrle}, \citenamefont {Ferr{\'e}}, \citenamefont {N{\`y}vlt},\ and\ \citenamefont {Vi{\v{s}}{\v{n}}ovsk{\`y}}}]{Hamrle2002}%
  \BibitemOpen
  \bibfield  {author} {\bibinfo {author} {\bibfnamefont {J.}~\bibnamefont {Hamrle}}, \bibinfo {author} {\bibfnamefont {J.}~\bibnamefont {Ferr{\'e}}}, \bibinfo {author} {\bibfnamefont {M.}~\bibnamefont {N{\`y}vlt}},\ and\ \bibinfo {author} {\bibfnamefont {{\v{S}}.}~\bibnamefont {Vi{\v{s}}{\v{n}}ovsk{\`y}}},\ }\bibfield  {title} {\bibinfo {title} {In-depth resolution of the magneto-optical kerr effect in ferromagnetic multilayers},\ }\href@noop {} {\bibfield  {journal} {\bibinfo  {journal} {Phys. Rev. B}\ }\textbf {\bibinfo {volume} {66}},\ \bibinfo {pages} {224423} (\bibinfo {year} {2002})}\BibitemShut {NoStop}%
\bibitem [{\citenamefont {Straub}\ \emph {et~al.}(1996)\citenamefont {Straub}, \citenamefont {Vollmer},\ and\ \citenamefont {Kirschner}}]{Straub1996}%
  \BibitemOpen
  \bibfield  {author} {\bibinfo {author} {\bibfnamefont {M.}~\bibnamefont {Straub}}, \bibinfo {author} {\bibfnamefont {R.}~\bibnamefont {Vollmer}},\ and\ \bibinfo {author} {\bibfnamefont {J.}~\bibnamefont {Kirschner}},\ }\bibfield  {title} {\bibinfo {title} {Surface magnetism of ultrathin $\gamma$-fe films investigated by nonlinear magneto-optical kerr effect},\ }\href@noop {} {\bibfield  {journal} {\bibinfo  {journal} {Phys. Rev. Lett.}\ }\textbf {\bibinfo {volume} {77}},\ \bibinfo {pages} {743} (\bibinfo {year} {1996})}\BibitemShut {NoStop}%
\bibitem [{\citenamefont {Wenzel}\ \emph {et~al.}(1997)\citenamefont {Wenzel}, \citenamefont {Hubert},\ and\ \citenamefont {Kambersk{\`y}}}]{Wenzel1997}%
  \BibitemOpen
  \bibfield  {author} {\bibinfo {author} {\bibfnamefont {L.}~\bibnamefont {Wenzel}}, \bibinfo {author} {\bibfnamefont {A.}~\bibnamefont {Hubert}},\ and\ \bibinfo {author} {\bibfnamefont {V.}~\bibnamefont {Kambersk{\`y}}},\ }\bibfield  {title} {\bibinfo {title} {Transparent simulation of magneto-optic diffraction from arbitrary magnetic multilayers},\ }\href@noop {} {\bibfield  {journal} {\bibinfo  {journal} {J. Magn. Magn. Mater.}\ }\textbf {\bibinfo {volume} {175}},\ \bibinfo {pages} {205} (\bibinfo {year} {1997})}\BibitemShut {NoStop}%
\bibitem [{\citenamefont {Herrmann}\ \emph {et~al.}(2006)\citenamefont {Herrmann}, \citenamefont {L{\"u}dge}, \citenamefont {Richter}, \citenamefont {Georgarakis}, \citenamefont {Poulopoulos}, \citenamefont {N{\"u}nthel}, \citenamefont {Lindner}, \citenamefont {Wahl},\ and\ \citenamefont {Esser}}]{herrmann2006}%
  \BibitemOpen
  \bibfield  {author} {\bibinfo {author} {\bibfnamefont {T.}~\bibnamefont {Herrmann}}, \bibinfo {author} {\bibfnamefont {K.}~\bibnamefont {L{\"u}dge}}, \bibinfo {author} {\bibfnamefont {W.}~\bibnamefont {Richter}}, \bibinfo {author} {\bibfnamefont {K.}~\bibnamefont {Georgarakis}}, \bibinfo {author} {\bibfnamefont {P.}~\bibnamefont {Poulopoulos}}, \bibinfo {author} {\bibfnamefont {R.}~\bibnamefont {N{\"u}nthel}}, \bibinfo {author} {\bibfnamefont {J.}~\bibnamefont {Lindner}}, \bibinfo {author} {\bibfnamefont {M.}~\bibnamefont {Wahl}},\ and\ \bibinfo {author} {\bibfnamefont {N.}~\bibnamefont {Esser}},\ }\bibfield  {title} {\bibinfo {title} {Optical anisotropy and magneto-optical properties of ni on preoxidized cu (110)},\ }\href@noop {} {\bibfield  {journal} {\bibinfo  {journal} {Phys. Rev. B}\ }\textbf {\bibinfo {volume} {73}},\ \bibinfo {pages} {134408} (\bibinfo {year} {2006})}\BibitemShut {NoStop}%
\bibitem [{\citenamefont {Keay}\ and\ \citenamefont {Lissberger}(1968)}]{Keay1968}%
  \BibitemOpen
  \bibfield  {author} {\bibinfo {author} {\bibfnamefont {D.}~\bibnamefont {Keay}}\ and\ \bibinfo {author} {\bibfnamefont {P.}~\bibnamefont {Lissberger}},\ }\bibfield  {title} {\bibinfo {title} {Longitudinal kerr magneto-optic effect in multilayer structures of dielectric and magnetic films},\ }\href@noop {} {\bibfield  {journal} {\bibinfo  {journal} {Optica Acta: Int J Opt}\ }\textbf {\bibinfo {volume} {15}},\ \bibinfo {pages} {373} (\bibinfo {year} {1968})}\BibitemShut {NoStop}%
\bibitem [{\citenamefont {Werner}\ \emph {et~al.}(2009)\citenamefont {Werner}, \citenamefont {Glantschnig},\ and\ \citenamefont {Ambrosch-Draxl}}]{Werner2009}%
  \BibitemOpen
  \bibfield  {author} {\bibinfo {author} {\bibfnamefont {W.~S.}\ \bibnamefont {Werner}}, \bibinfo {author} {\bibfnamefont {K.}~\bibnamefont {Glantschnig}},\ and\ \bibinfo {author} {\bibfnamefont {C.}~\bibnamefont {Ambrosch-Draxl}},\ }\bibfield  {title} {\bibinfo {title} {Optical constants and inelastic electron-scattering data for 17 elemental metals},\ }\href@noop {} {\bibfield  {journal} {\bibinfo  {journal} {Journal of Physical and Chemical Reference Data}\ }\textbf {\bibinfo {volume} {38}},\ \bibinfo {pages} {1013} (\bibinfo {year} {2009})}\BibitemShut {NoStop}%
\end{thebibliography}%

\onecolumngrid
\section{Supplementary Materials}
\label{sec Appendix}

In this supplementary materials we provide a tabulation of the material parameters and simulation conditions. We also demonstrate the probe-angle dependence of Complex Kerr-response.
\subsection{Simulation parameters }
\begin{table}[H]
\begin{center}
\begin{tabular}{||c c c||} 
 \hline
 Parameter & Value & \\ [0.5ex] 
   \hline
   Voigt Constant & $ (1 + 8i) \times 10^{-3} $ \zit{herrmann2006} & \\
 Probe wavelength $\lambda_{\text{probe}}$   & $\SI{400}{\nano\meter}$  &\\
 Absorbed fluence $F_{0}$ (diffusive transport)  & $\SI{200}{\joule{\meter}^{2}}$  &\\
 Absorbed fluence $F_{0}$ (ballistic transport)  & $\SI{62}{\joule\meter^{2}}$   &\\  
 FWHM (diffusive transport) &  $\SI{50}{\femto\second}$ &\\ 
 FWHM (ballistic transport)  & $\SI{50}{\femto\second}$  &\\ 
 Penetration depth (diffusive transport)  & $\SI{15}{\nano\meter}$  &\\ 
 Penetration depth (ballistic transport)  & $\infty$ &\\
 Thickness $L$  & $\SI{40}{\nano\meter}$  &\\ 
 Transport coefficients  & \zit{Ashok2022}  &\\ 
   \hline 
  \hline 
\end{tabular}
\end{center}
\caption{\label{table: sim_vals} Table of parameters and their numerical values used to simulate angle-dependent Complex-MOKE dynamics.}
\end{table}
%\subsection{Experimental parameters}
%WIP
%\newpage

\subsection{Complex Kerr-sensitivity function}

The probe-angle dependent Complex Kerr-response dynamics is due to the confluence of depth-dependence of Complex Kerr-sensitivity  and time- and depth-dependent magnetization.
In order to demonstrate the probe-angle dependence, we consider a \textit{snapshot} of Kerr-response at \SI{500}{fs} and demonstrate that the magnitude of Kerr-rotation is lower than the initial value for all probe-angles where as the magnitude Kerr-ellipticity is probe-angle dependent. 
\begin{figure}[H]
    \centering
     \hspace*{-.5cm}
    \includegraphics[width=\textwidth]{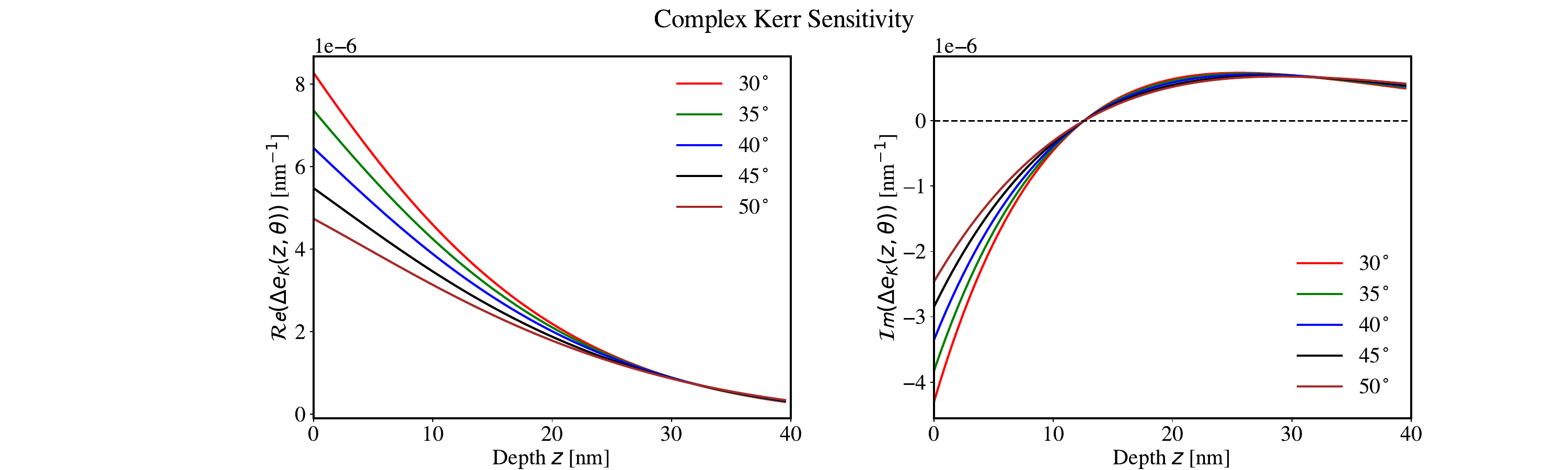}
    %need to figure out how to insert large dpi figures into latex
    \caption{The real and imaginary parts of the complex Kerr sensitivity $\Delta e_{\text{K}}(z, \theta_{i})$ are plotted as a function of depth $z$ for various probe angles. The real part of the sensitivity $\mathcal{R}e(\Delta e_{\text{K}}(z))$ is positive and monotonically decreases with depth irrespective of the probe angle $\theta_{i}$. The imaginary part of the Kerr sensitivity $\mathcal{I}m(\Delta e_{\text{K}}(z))$ is quantitatively dependent on the probe angle. While it has a negative magnitude closer to the surface, it monotonically increases with depth up-to around penetration depth of the probe beam. Beyond which it monotonically decreases and reaches negligible but positive value.}
    \label{fig:cmplx_kerr_sensitivity}
\end{figure}

\subsection{Interplay of spatially dependent complex Kerr-sensitivity and spatially dependent magnetization dynamics}
\vspace{0.00mm} 

\begin{figure}[ht]
    \centering
     \hspace*{-.5cm}
    \includegraphics[width=\textwidth]{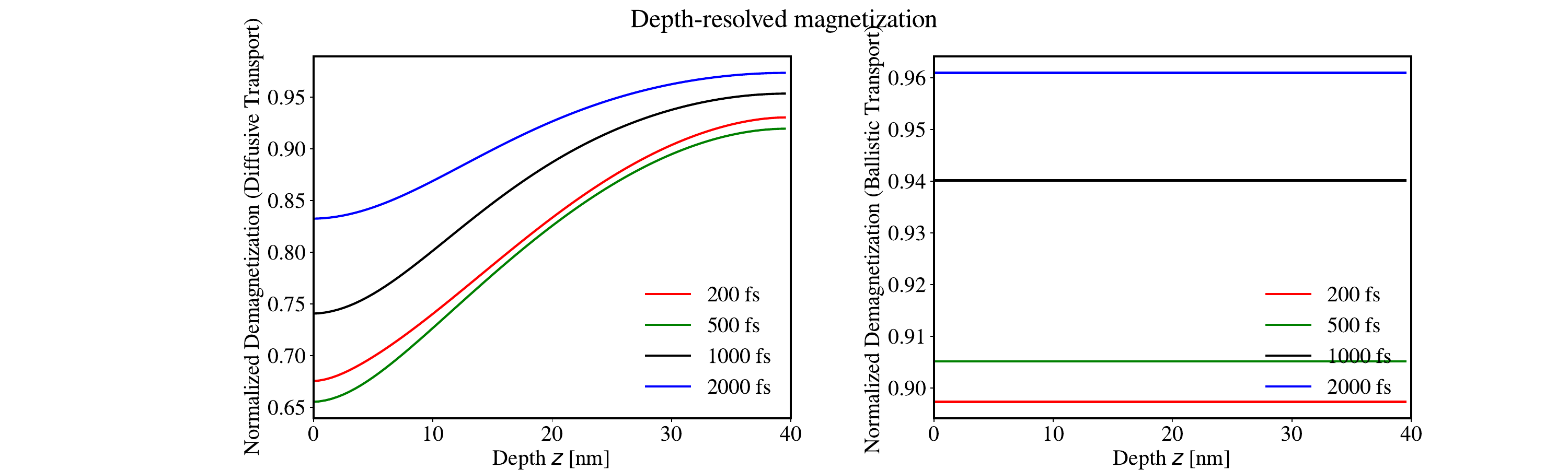}
    %need to figure out how to insert large dpi figures into latex
    \caption{Normalized magnetization $m(z)$ is plotted as a function of depth $z$ at various time instances of $\SI{200}{\femto\second}, \SI{500}{\femto\second}, \SI{1000}{\femto\second} \text{ and } \SI{2000}{\femto\second}$ after laser excitation. Diffusive and Ballistic transport cases are separately shown. In the presence of diffusive transport only the magnetization dynamics is inhomogeneous. This inhomogeneity is strongly dependent on time as well as depth. In contrast, when ballistic transport is predominant, the inhomogeneities in magnetization are rapidly washed out. This leads to a depth independent magnetization irrespective of the time.}
    \label{fig:depth_reslvd_magnetization}
\end{figure}
\vspace{0.00mm} 
%\newpage
\begin{figure}[ht]
    \centering
     \hspace*{-.5cm}
    \includegraphics[width=\textwidth]{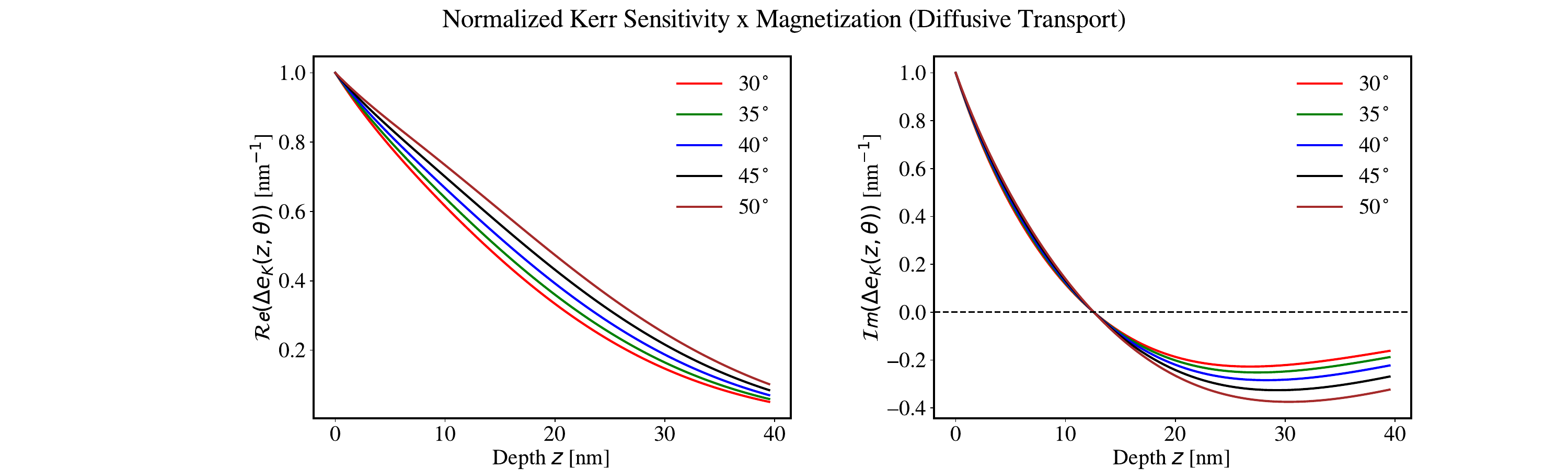}
    %need to figure out how to insert large dpi figures into latex
    \caption{Depth dependent complex Kerr sensitivity is weighted by the depth dependent magnetization at 500 fs $\Delta e_{\text{K}}(z, \theta_{i}) \times m(z)$ and is plotted as a function of depth $z$. Here diffusive transport is the predominant transport mechanism (Fig.[\ref{fig:depth_reslvd_magnetization}]). The data is normalized to the it's value at the Surface ($z=0$). Figure shows the influence of inhomogeneities of magnetization present due the diffusive transport on the differential complex Kerr-signal. The real part is only marginally influenced by the inhomogeneities in magnetization at depths lower than the information depth of the probe pulse irrespective of the probe angle. The strongest contribution comes from regions closer to the surface and is negative for all depths $z > 0$. The contributions for the imaginary part is strongly dependent on probe angle as well as depth. For certain probe angles, the \textit{positive} regions have a stronger contribution compared to the \textit{negative} regions, leading to an increase in relative Kerr-ellipticity. The opposite holds true for probe angles lower than $40^{o}$. This is made clear in Fig.[\ref{fig:cumulative_signal_diffusive}].}
    \label{fig:magxsensdiff}
\end{figure}
\vspace{0.00mm} 

\begin{figure}[ht]
    \centering
     \hspace*{-.5cm}
    \includegraphics[width=\textwidth]{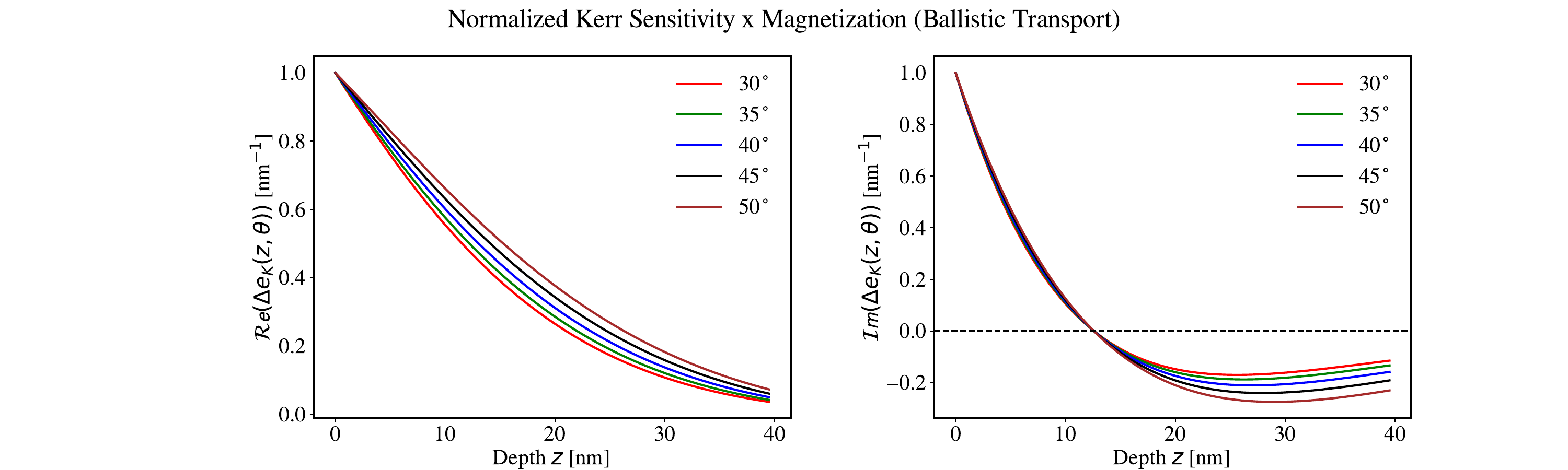}
    %need to figure out how to insert large dpi figures into latex
    \caption{Depth dependent complex Kerr sensitivity is weighted by the depth dependent magnetization $\Delta e_{\text{K}}(z, \theta_{i}) \times m(z)$ at 500 fs with the ballistic transport is the predominant transport mechanism (Fig.[\ref{fig:depth_reslvd_magnetization}]). Figure shows the influence of homogeneous magnetization in the ballistic case on the differential complex Kerr-signal. The real part is weakly influenced by the magnetization at depths lower than the information depth of the probe pulse irrespective of the probe angle. The contributions for the imaginary part is strongly dependent on probe angle as well as depth. For certain probe angles, regions closer to the surface contributes positively to the change in degree of ellipticity.}
    \label{fig:magxsensball}
\end{figure}
\vspace{0.00mm} 
%\newpage
\begin{figure}[ht]
    \centering
     \hspace*{-.5cm}
    \includegraphics[width=\textwidth]{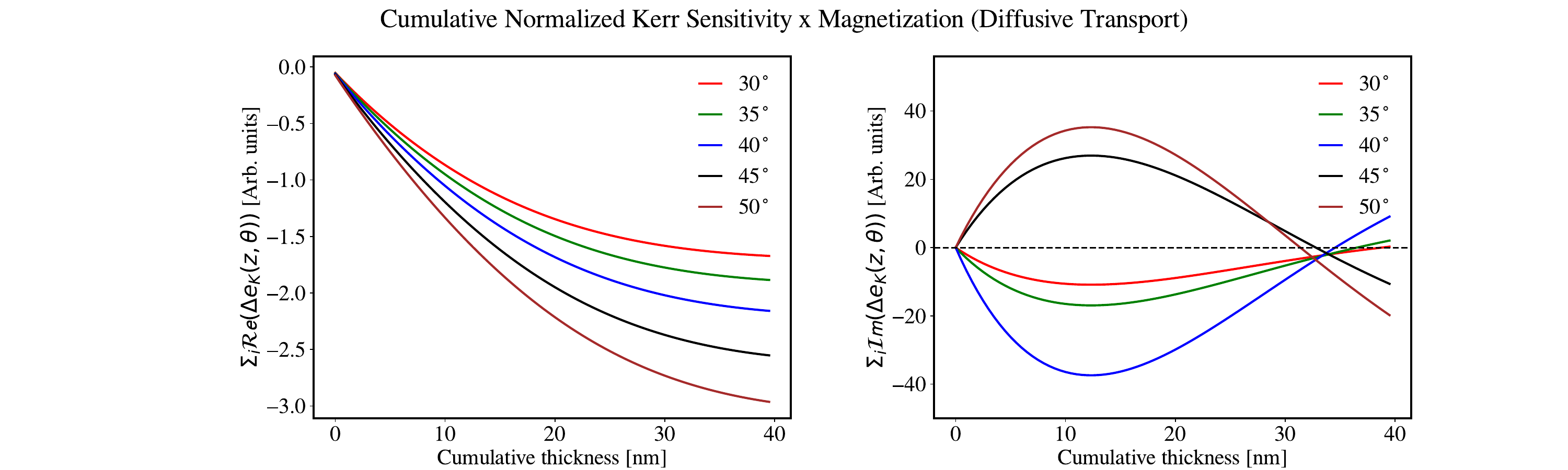}
    %need to figure out how to insert large dpi figures into latex
    \caption{Cumulative signal (Integrated Kerr-response up to a given depth) at the given depth is plotted as a function of depth for the diffusive transport case at 500 fs.  The numerical values are relative to the total signal at t=0 for various probe angles . The real part is weakly dependent on the probe angle and the strongest cumulative contribution comes from depths lower than the information depth. It's cumulative response is always lower than it's value at time $t = 0$ irrespective of the probe angle. The imaginary part of the cumulative signal is strongly dependent on the probe angle as well as the depth. The total signal (cumulative signal at $z = \SI{40}{\nano\meter}$) is positive for some angles while being negative for other probe angles. This effectively leads to the time-dependent increase of Kerr-ellipticity for some probe angles.}
    \label{fig:cumulative_signal_diffusive}
\end{figure}
\begin{figure}[h]
    \centering
     \hspace*{-.5cm}
    \includegraphics[width=\textwidth]{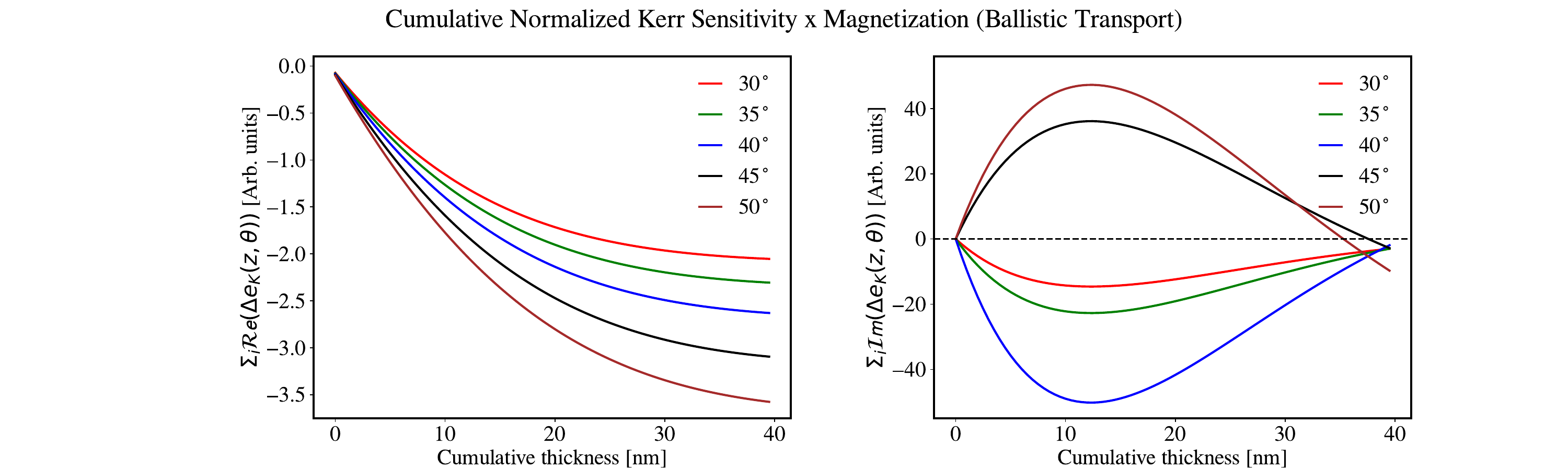}
    \caption{ Cumulative signal (Integrated Kerr-response up to a given depth) at the given depth is plotted as a function of depth for the ballistic transport case at 500 fs. The numerical values are relative to the total signal at $t = 0$ for various probe angles. The real part is weakly dependent on the probe angle and the strongest cumulative contribution comes from depths lower than the information depth. It's cumulative response is always lower than it's value at time $t = 0$ irrespective of the probe angle. The imaginary part of the cumulative signal is strongly dependent on the probe angle as well as the depth. The total signal (cumulative signal at $z = \SI{40}{\nano\meter}$) is lower than the initial value for all probe angles. This effectively leads to the time-dependent decrease in Kerr-ellipticity dynamics.    }
    \label{fig:cumulative_signal_ballistic}
\end{figure}
\clearpage
\subsection{ \SI{10}{\nano\meter} Nickel}
To confirm our hypothesis that the magnetization gradients lead to probe-angle dependent Kerr-ellipticity dynamics where as when magnetization gradients are absent probe-angle dependence of Kerr-ellipticity dynamics vanishes, we performed a set of Kerr-ellipticity dynamics measurements and simulations for a 10 nm thick Nickel film.
\subsection{Experimental measurements of probe angle dependent Kerr dynamics for a \SI{10}{\nano\meter} thick Nickel film.}

\begin{figure}[htb]
    \centering
    \includegraphics[width=0.5\textwidth]{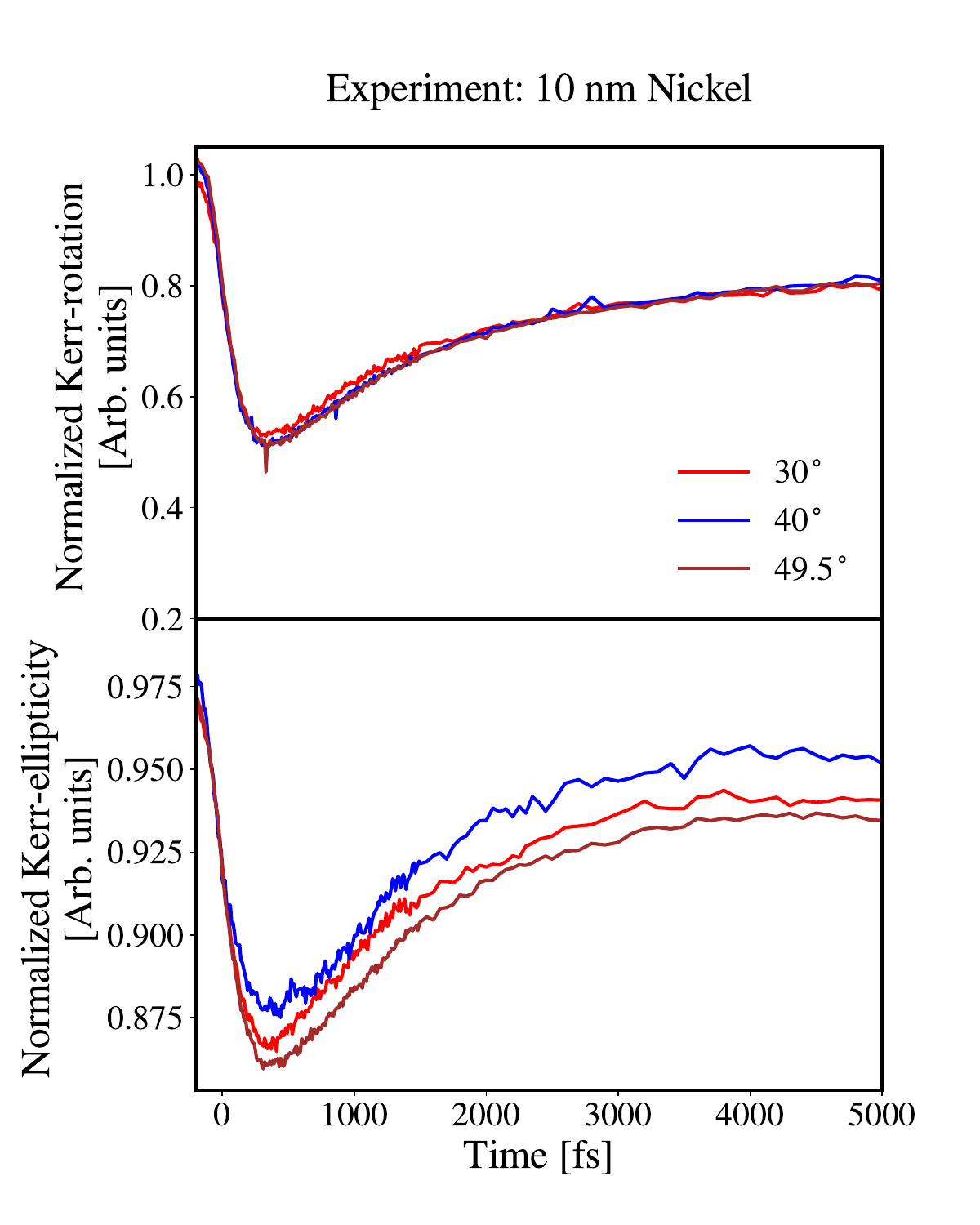}~

    \caption{Experimental results for Kerr-Rotation and -Ellipticity Dynamics measured for various probe angles for a \SI{10}{\nano\meter} film.  Comparing with the theoretical expectations, we conclude that the thin \SI{10}{\nano\meter}-film is homogeneously heated and magnetized. Therefore, the absence of gradients lead to probe-angle independence Kerr-rotation and -ellipticity dynamics.}
    \label{fig:results exp appendix}
\end{figure}

%\newpage 
\subsection{Theoretical simulations of probe angle dependent Kerr dynamics for a \SI{10}{\nano\meter} thick Nickel film.}
\begin{figure}[ht]
    \centering
    \includegraphics[width=0.5\textwidth]{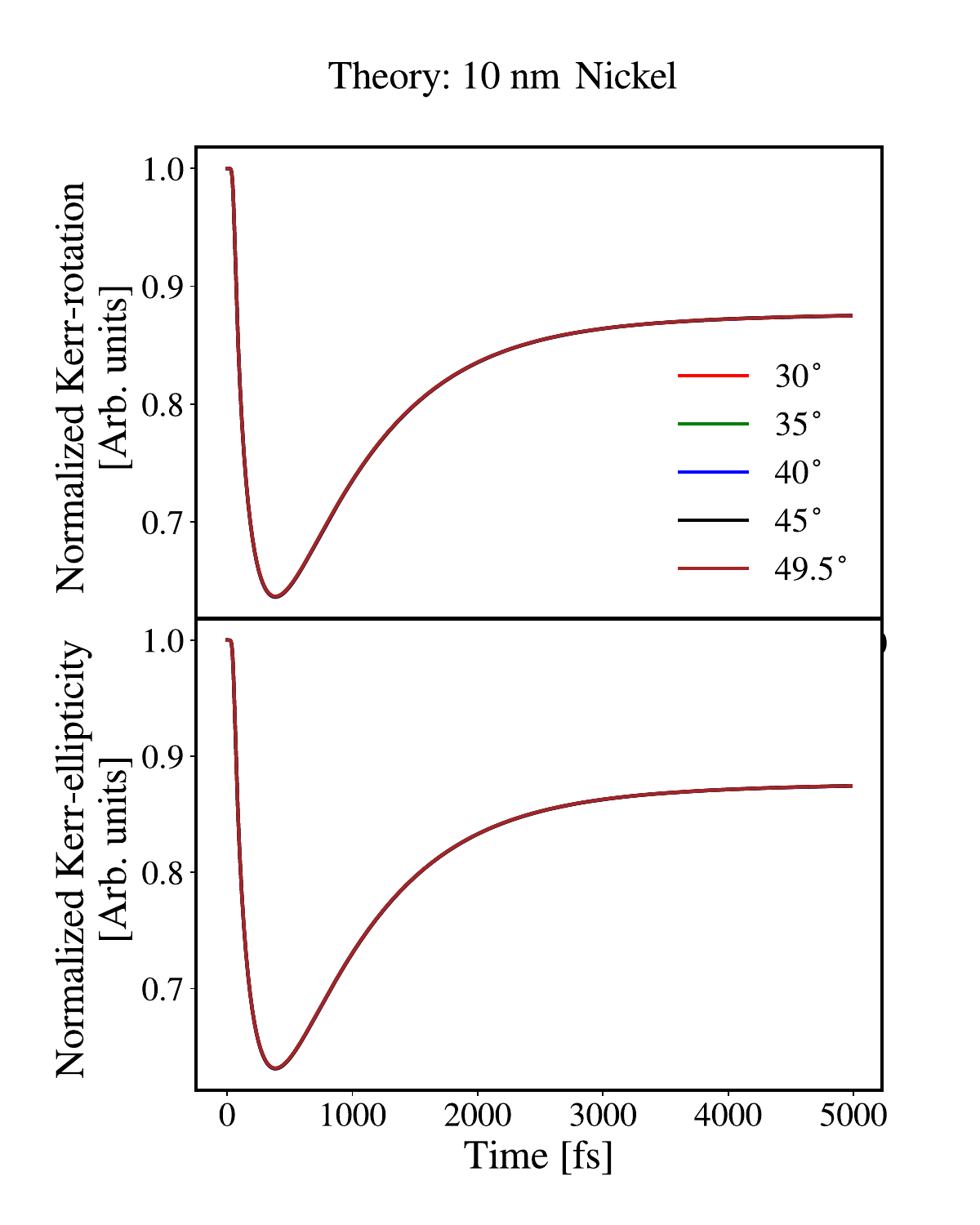}
    \caption{Theoretical results for Kerr-rotation and -ellipticity dynamics calculated for various probe angles for a \SI{10}{\nano\meter} film. A pump penetration depth of \SI{15}{\nano\meter} along with spin-resolved charge and heat transport erases any small gradients in magnetization.}
    \label{fig:results exp2}
\end{figure}
\end{document}